\documentclass[aps,reprint,prb]{revtex4-1}
\usepackage{graphicx}
\usepackage{dcolumn}
\usepackage{bm}
\usepackage{amsmath}
\usepackage{siunitx}
\usepackage{float}
\usepackage{gensymb}

\begin{document}

\title{Vortex Dynamics and Dissipation Under High-amplitude Microwave Drive} 

\author{Mattia~Checchin}
\email[]{checchin@fnal.gov}
\author{A.~Grassellino}
\affiliation{Fermi National Accelerator Laboratory, Batavia, Illinois 60510, USA}

\date{\today}

\begin{abstract}In this paper, we describe the vortex dynamics under high-amplitude microwave drive and its effect on the surface resistance of superconductors. The vortex surface resistance is calculated with a Montecarlo approach, where the vortex motion equation is solved for a collection of vortex flux lines each oscillating within a random pinning landscape. This approach is capable of providing a detailed description of the microscopic vortex dynamics and in turn important insights into the microwave field amplitude dependence of the vortex surface resistance. The numerical simulations are compared against experimental data of vortex surface resistance at high microwave amplitude measured by means of bulk niobium superconducting-radio frequency cavities operating at 1.3~GHz. The good qualitative agreement of simulations and experiments suggests that the non-linear dependence of the trapped flux surface resistance with the microwave field amplitude is generated by progressive microwave depinning and vortex jumps.
\end{abstract}


\maketitle 

\section{Introduction}
Upon cooldown below critical temperature in presence of an external magnetic field, magnetic flux quanta\textemdash so-called vortexes\textemdash can exist in thermodynamic equilibrium in the mixed state of type-II superconductors.\cite{Abrikosov_ZETF_1957, Shubnikov_ZETF_1937}

Below the lower critical field vortexes are not stable in the superconductor, however, because of the occurrence of defects in real materials, vortexes get pinned and survive even in the Meissner state. Incomplete Meissner effect where 100\% of the applied field is trapped by the superconductor can be achieved with superconducting radio-frequency (SRF) cavities by field-cooling  slowly across transition, as experimentally observed by many studies.\cite{Romanenko_JAP_2014,Romanenko_APL_2014_2,Posen_JAP_2016,Martinello_APL_2016}

Driven by radio-frequency (RF) and microwave fields, vortexes oscillate dissipating power and contributing to the total surface resistance. Other than vortex oscillation, others contributions to the surface resistance are: the temperature dependent surface resistance due to thermally-excited quasi-particles above the gap,\cite{Mattis_PhysRev_1958} surface resistance due to sub-gap states,\cite{Gurevich_SUST_2017} proximity-coupled normal-conducting inclusions (or oxides),\cite{Kubo_PRB_2019} nano-hydrides,\cite{Romanenko_SUST_2013} and dielectric dissipation due to spurious two-level systems at low microwave amplitudes and low temperatures.\cite{Romanenko_PRL_2017,Romanenko_arXiv_2018}
\begin{figure*}[t]
\centering
\includegraphics[width=17cm]{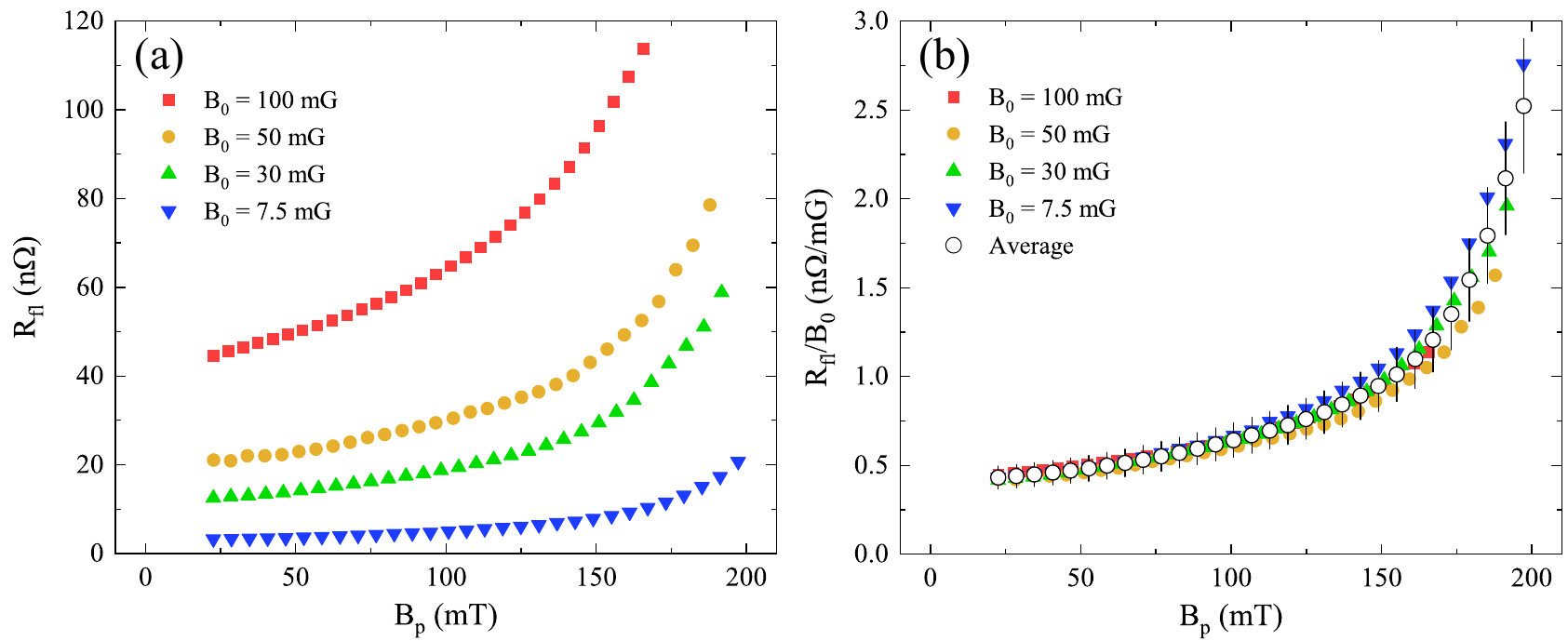}
\caption{In graph (a), the experimental data of trapped flux surface resistance up to high values of peak magnetic field is reported, while the normalized $R_{fl}$ over the trapped magnetic field is reported in graph (b).}
\label{fig.resistance}
\end{figure*}

Vortex dissipation under microwave drive is a topic that has been under investigation for many decades. Initial studies on type-II superconducting films\cite{Cadorna_PRL_1964} brought about the development of models to describe the vortex surface resistance dependence on frequency.\cite{Gittleman_PRL_1966,Rabinowitz_JAP_1971} Since then, several other studies investigated the microwave behavior of conventional superconductors,\cite{Chin_PRB_1992,Janjusevic_PRB_2006,Alimenti_IEEE_2019} cuprates,\cite{Revenaz_PRB_1994,Matsuda_PRL_1995,Golosovsky_SUST_1996} and iron-based superconductors\cite{Hashimoto_PRL_2009_1,Hashimoto_PRL_2009_2, Okada_PRB_2012} in presence of vortexes. In parallel, many theoretical works allowed to describe the expected penetration depth change due to vortex oscillation,\cite{Coffey_PRL_1991} the vortex surface impedance temperature-dependence in high-$\text{T}_\text{c}$ superconductors,\cite{Marcon_PRB_1991} and the precise electrodynamics of vortexes lattices in the mixed state.\cite{Sonin_PRB_1992}

In the framework of modern particle accelerators employing SRF cavities as accelerating devices, dissipation due to vortex motion under microwave drive is a topic of central importance, inasmuch trapped-fields in the order of $\sim$1\% of the earth magnetic field can increase the surface resistance of about 1~n$\Omega$,\cite{Martinello_APL_2016} a large value for state-of-the-art cavities capable of achieving surface resistance values as low as 4~n$\Omega$ at 2~K.\cite{Grassellino_SUST_2013,Martinello_PRL_2018}

Early studies on S-band SRF bulk Nb cavities\cite{Piosczyk_IEEE_1973} recognized the effect of vortexes on the surface resistance, identifying the microwave amplitude dependence of the latter to follow a non-linear behavior. Later studies showed that up to moderate gradients the microwave field amplitude dependence of the vortex surface resistance in Nb-sputtered on copper SRF cavities is linear.\cite{Benvenuti_PhysicaC_1999} Studies of vortex dynamics in cavity-grade Nb and theoretical description of the phenomenon observed in the zero microwave field amplitude approximation were also studied in detail.\cite{Gurevich_PRB_2013}
 
More recently, intensive studies of the trapped flux surface resistance at low and moderate accelerating gradients as a function of the electron mean-free-path ($l$) discovered a bell-like trend as a function of $l$,\cite{Martinello_APL_2016, Gonnella_JAP_2016} which is well described by the interplay of pinning- and flux-flow-limited dissipation.\cite{Checchin_SUST_2017,Checchin_APL_2018, Calatroni_IEEE_2017} The studies also reported an almost linear dependence of the vortex dissipation with the microwave field amplitude in the resonator up to moderate values.

The microwave field amplitude dependence can be understood as the occurrence of non-quadratic terms in the pinning potential.\cite{Calatroni_PRAB_2019} The linear dependence was shown to be well described by adding a cubic term to a parabolic pinning potential in the vortex motion equation.\cite{Calatroni_PRAB_2019} However, this approach is limited since built around the single-vortex dynamics, while in the material there are several vortexes contributing to the overall dissipation, each one interacting with its surrounding.

A second approach,\cite{Liarte_PRApplied_2018} that overcomes the latter issue, implements a mean-field method to solve the vortex motion equation for collective weak pinning. In this case the microwave field amplitude dependence has two regimes: i) hysteretic losses at low amplitudes where the dependence is linear, and ii) viscous losses at higher amplitudes where the surface resistance saturates to a constant value.

In the present work, the high-amplitude microwave field dependence of the vortex surface resistance is explored by comparing the experimental data acquired for a high-performing single cell Nb accelerating cavity to the surface resistance calculated by means of numerical simulations. The model developed is based on a Montecarlo approach that assumes a random distribution of pinning potentials in the vortex oscillation plane. The simulation scheme here presented allows to gather insights into the dynamics of vortexes oscillation in random pinning potentials such us vortex jumps an microwave depinning, which are not captured by previous models, but key to correctly interpret the experimental data.

\section{Experimental Data\label{sec.data}}
Experimental data of vortex surface resistance was estimated from microwave measurements of a single-cell TESLA-type\cite{TESLA_Cavities_PRST_2000} bulk niobium cavity operating at 1.3~GHz.

The cavity surface was prepared with a series of treatments designed to increase the maximum accelerating gradient achievable by the resonator, so that to capture the power dissipation fingerprint due to the high-amplitude microwave behavior of vortexes. The cavity was initially electropolished to remove 200~$\mu$m and subsequently baked at \SI{800}{\degreeCelsius} for 3 hours to degas hydrogen and release internal stress. After this initial surface preparation the cavity was high-pressure water rinsed in a class 10 clean-room to remove field-emitting particles form the inner surface and assembled with antennae needed for microwave characterization. The cavity was then evacuated and baked at \SI{120}{\degreeCelsius} for 48 hours in situ.

The resonator was equipped with three single-axis flux gates magnetometers from Bartington Instruments located symmetrically around the cavity equator, a set of Helmholtz coils aligned coaxially to the cavity beam pipe, and a temperature mapping system (Tmap)\cite{Knobloch_T_Map_RSI_1994} to precisely monitor the cavity temperature rise during the measurements.

The cavity was initially zero-field-cooled (ZFC) to 1.5~K and data Q-factor versus accelerating gradient acquired. In such conditions, the resonator reached maximum peak magnetic field ($B_p$) of 204.5~mT (48~MV/m) with Q-factor of $9.6\cdot10^9$. The same cavity was then field-cooled (FC) very slowly to 1.5~K four times trapping respectively 7.5~mG, 30~mG, 50~mG, and 100~mG upon the superconducting transition.

As described in detail in Ref.~\onlinecite{Martinello_APL_2016}, the vortex surface resistance is calculated by subtracting ZFC data from FC data acquired at the same temperature. By performing the measurement at 1.5~K, the thermal contribution to the surface resistance is negligible, and the systematic error introduced by the accuracy of the temperature control in the dewar minimized.

In Fig.~\ref{fig.resistance}(a), the vortex surface resistance ($R_{fl}$) as a function of the peak magnetic field is reported. The normalized vortex surface resistance over the magnetic field trapped during the cooldown ($B_0$)\textemdash so-called trapped-flux sensitivity\textemdash is reported in Fig.~\ref{fig.resistance}(b).
\begin{figure*}[t]
\centering
\includegraphics[width=17cm]{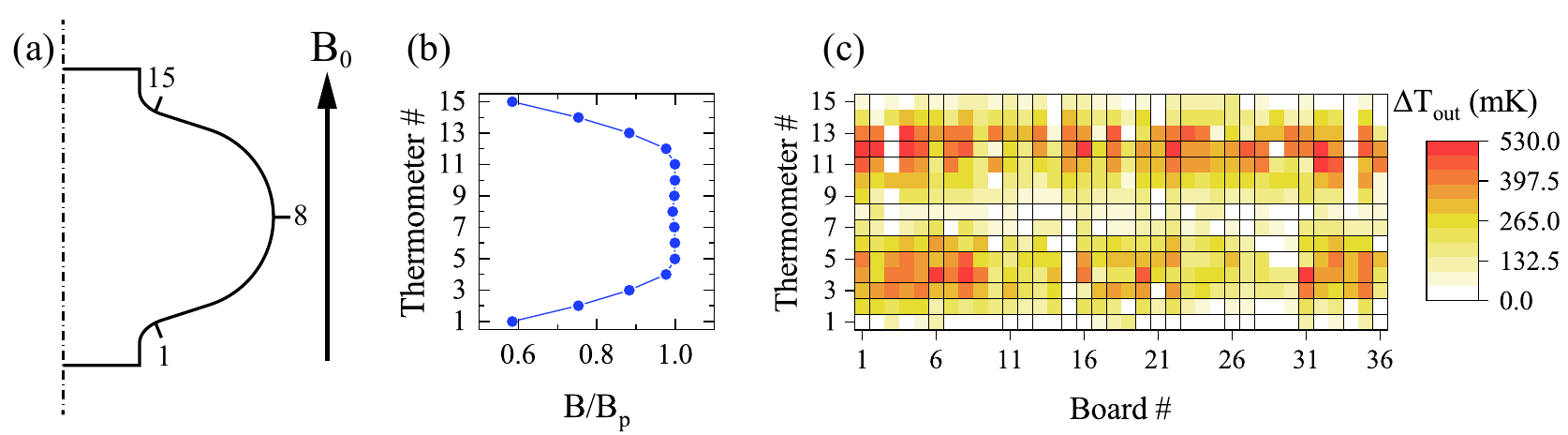}
\caption{A sketch of the cavity section indicating the location of thermometers in shown in figure (a). Graph (b) shows the ratio between local magnetic field at the surface over the peak magnetic field at each thermometer location, while the Tmap showing vortex dissipation collected at $B_p=166$~mT is reported in (c).}
\label{fig.Tmap}
\end{figure*}

As shown in Fig.~\ref{fig.resistance}(b), once the vortex surface resistance is normalized by $B_0$, the field dependence appears the same within the measurement error ($\sim$10\% of $R_{fl}$ value)\cite{Melnychuk_RevSciInstr_2014,Martinello_APL_2016}. This means that, within the range of $B_0$ and $B_p$ investigated, the peak magnetic field amplitude dependence of $R_{fl}$ is dictated by vortex dynamics and not by thermal feedback.

In agreement with previous measurement of vortex surface resistance as a function of accelerating gradient performed in thin film and bulk Nb cavities,\cite{Benvenuti_PhysicaC_1999,Martinello_APL_2016} $R_{fl}$ increases almost linearly up to moderate fields ($\sim$80~mT). For higher field amplitudes the vortex surface resistance deviates from the almost linear growth and the slope increases with $B_p$, leading to higher values of resistance.

Experimental data was also acquired by means of the Tmap system, composed of 36 boards arranged symmetrically around the cavity, each incorporating 15 thermometers in contact with the outer surface of the cavity as sketched in Fig.~\ref{fig.Tmap}(a).

The data was acquired at $B_p=166$~mT both in the ZFC experiment and in the FC experiment with 100~mG applied during the cooldown. The ZFC tmap data was then subtracted from the FC data in order to obtain the heating pattern due to trapped-vortexes only. Temperature mapping data is shown in Fig.~\ref{fig.Tmap}(c). The board number is reported in the abscissa, while the thermometer number along the ordinate.

Interestingly, the vortex dissipation shows a specific pattern with heating concentrated symmetrically with respect to the equator of the cavity (the equator location corresponds to thermometer number 8), where, instead, the temperature differential $\Delta T_{out}$ is much lower than neighboring areas and tends to zero.

In this configuration, the trapped magnetic field $B_0$ is parallel to the cavity beam axis (as shown in Fig.~\ref{fig.Tmap}(a)) and at the equator the microwave currents are parallel to the trapped field. Because of the angular dependence of the Lorentz force that drives vortex oscillation, this geometrical effect leads to lower dissipation at the equator compared to areas of the cavity where the flux has an angle almost perpendicular to the microwave currents.

Important to point out that this behavior is intertwined with the modulation of the microwave currents that varies with the local amplitude of the surface magnetic field $B$ at the cavity surface. $B$ is almost constant from thermometer 4 to thermometer 12\textemdash surface peak field $B_p$ at thermometer 6 and 10, and it decays rapidly to zero from thermometer 3 to 1 and from 13 to 15 (as shown in Fig.~\ref{fig.Tmap}(b)).

This modulation of the surface field $B$ explains why the vortex dissipation is almost zero at the cavity iris (thermometers 1 and 15) where flux and currents are instead close to orthogonality.

\section{Vortex Dynamics}
Numerical simulations of vortex dynamics were performed in order to shed light on the field dependence of the vortex surface resistance.

Let's assume a superconducting surface at $z=0$ parallel to the plane $xy$, where the semi-space $z>0$ is filled by superconducting niobium and the semi-space $z<0$ is vacuum free space. The magnetic field trapped ($B_0$) is parallel to the $z$ axis, and we assume that upon transition the magnetic flux is subdivided in individual vortexes not interacting with each other $B_0<<B_{c2}$.

The microwave excitation\textemdash with frequency $\omega/2\pi$\textemdash at the superconducting surface has magnetic field oscillating along the $x$ axis, inducing current parallel to the $y$ axis, which decays exponentially in the material along $z$, with characteristic decay length equal to the penetration depth $\lambda$.\cite{London_ProcRSocLond_1935} The amplitude of the microwave field coincides with the peak magnetic field $B_p$.

Neglecting the inertial term, the vortex motion equation has form:
\begin{equation}
\label{eq.motion}
    \eta\Dot{u}\left( t,z\right)=\epsilon u''\left( t,z\right)+f_p\left( t,z\right)+\gamma B_p\cos\left(\omega t\right) e^{-z/\lambda}\text{,}
\end{equation}
where $u\left( t,z\right)$ is the vortex displacement as a function of time and depth $z$, $\Dot{u}$ the time partial derivative, $u''$ the second partial derivative along $z$, and $\gamma=\phi_0/(\mu_0\lambda)$. $\epsilon$ is the single-vortex line tension,\cite{Brandt_RepProgPhys_1995} defined as: 
\begin{equation}
  \epsilon=\dfrac{\phi_0^2}{4\pi\mu_0\lambda^2}\left[
\dfrac{1}{2}+\ln\kappa+e^{-0.4-0.8\ln\kappa-0.1\ln^2\kappa}
\right]\text{,}
\end{equation}
with $\kappa=\lambda/\xi$. The vortex viscosity $\eta$ is equal to $\sigma\phi_0 B_{c2}$,\cite{Bardeen_PR_1965} and $\sigma$ is the niobium normal-state conductivity at low temperature. The pinning force per unit of length is introduced through $f_p\left( t,z\right)$.

\subsection{Pinning Landscape}
Vortex pinning in superconductors is explained as the tendency of the system to minimize the total loss in condensation energy. Being a vortex itself a singularity in the condensation energy, it interacts with areas in the material where superconductivity is suppressed or absent in order to minimize the overall condensation energy loss.

This mechanism intrinsically implies the minimum dimension for a defect to be an efficient pinning center as the coherence length $\xi$, which represents the minimum length over which a given change in condensed electron density can be made.\cite{Ginzburg_ZETF_1950}

The coherence length in niobium is of the order of $38$~nm or lower when scattering centers are present. This implies that any defect in the material with dimension of the order of $\xi$ or larger could pin vortexes. Due to their small transverse dimension ($\sim1-5$~nm), grain boundaries and single dislocations cannot be consider efficient pinning centers, whereas dislocations tangles, and material non-uniformity can extend for even hundreds of nanometers, allowing for more efficient pinning.\cite{Martinello_SRF_2019,Antoine_PRAB_2019}

Due to the defective nature of real materials, many defects may pin simultaneously a single vortex along its length, making virtually impossible to measure directly the pinning potential, especially for bulk materials. Theoretical calculations\cite{Thuneberg_PRB_1984} suggests that isolated pinning centers can be approximated by Lorentzian functions of the type $U(r)=-U_0/(1+(r/\xi)^2)$, with $r$ the radial distance form the vortex on the plane orthogonal to $\phi_0\hat{n}$, while experimental measurements on thin films\cite{Embon_SRep_2015} showed that real pinning potentials can be approximated by a summation of several of these Lorentzian functions. 

In the case under study, the pinning landscape is defined in the oscillation plane of vortexes\textemdash $xz$, and it is constructed through a random bottom up approach based on the summation of $N$ Lorentzian pinning potentials described above:
\begin{equation}
\label{eq.pinning}
    \begin{split}
        &U(x,z)=\\
        &-\sum_i^N\Bigg[\dfrac{U_i (b\xi)^2}{(b\xi)^2+\big(x-X_i\big)^2+a_i\big(z-Z_i\big)^2}\Bigg]\text{.}
    \end{split}
\end{equation}

Here, $U_i$, $X_i$, and $Z_i$ represent respectively the $i$-th pinning potential energy and ($x$,$z$) coordinates in the vortex oscillation plane, the variable $a_i$ is a dimensionless parameter that accounts for the anisotropy of the pinning potential along the $z$ direction, while $b$ is a multiplicative factor to account for the pinning center dimension. The pinning force is defined by the summation of the negative derivative along $x$ of the $i$-th pinning potential, over the total number of pinning centers $N$.

The coordinates $(X_i, Z_i)$ and the anisotropy values $a_i$ are generated randomly within a certain domain, usually $-1<X_i<1$~$\mu$m, $0<Z_i<3$~$\mu$m, and $0.001<a_i<1$. The pinning energy values $U_i$ are generated randomly following a normal distribution centered around an average value ($U_{ave}$). $U_{ave}$ is defined as the pinning energy that translates to a maximum pinning force per unit of length ($f_{p,ave}$) comparable to values found in literature. Usually for niobium the pinning force per unit of length varies depending on the material impurity content and crystallographic structure in between $10^{-6}-10^{-4}$~N/m.~\cite{Allen_PRB_1989,Park_PRL_1992}

The input value $f_{p,ave}$, together with the pinning potential depth distribution normal deviation ($\sigma_U=5\%U_{ave}$), the coordinates $(X_i, Z_i)$, and parameters $a_i$ and $b$, describe completely the random pinning landscape used in the calculation.

\subsection{Numerical Simulations}
Numerical simulations of the vortex motion are defined within the $xz$ plane domain where the pinning force landscape is defined by means of Eq.~\ref{eq.pinning}, assuming distributions of pinning energy, location, and anisotropy as described above.

The initial condition and the boundary conditions used in the calculation are the following:
\begin{equation}
    \begin{cases}
        &u(0,z)=0\\
        &u'(t,0)=0\\ 
        &u'(t,z_{max})=0\text{,}
    \end{cases}
    \label{eq.BC}
\end{equation}
where $z_{max}$ represent the maximum depth of the solution domain defined from 0\textemdash the microwave surface, to $z_{max}$.

In several works,\cite{Gurevich_PRB_2013,Liarte_PRApplied_2018} the boundary condition at the far edge of the solution domain is set to be $u(t,z_{max})=0$, implying a vortex strongly pinned in the bulk. On the opposite, in this work vortexes are allowed to find their own equilibrium position for any randomly-generated pinning landscape and we set $u'(t,z_{max})=0$. As we will see later, this boundary conditions choice is mandatory to allow the vortex line displacement in the bulk to evolve with time until an equilibrium position is found, condition reached after several oscillations when the simulation converges to a steady-state solution.
\begin{figure}[b]
\centering
\includegraphics[width=8cm]{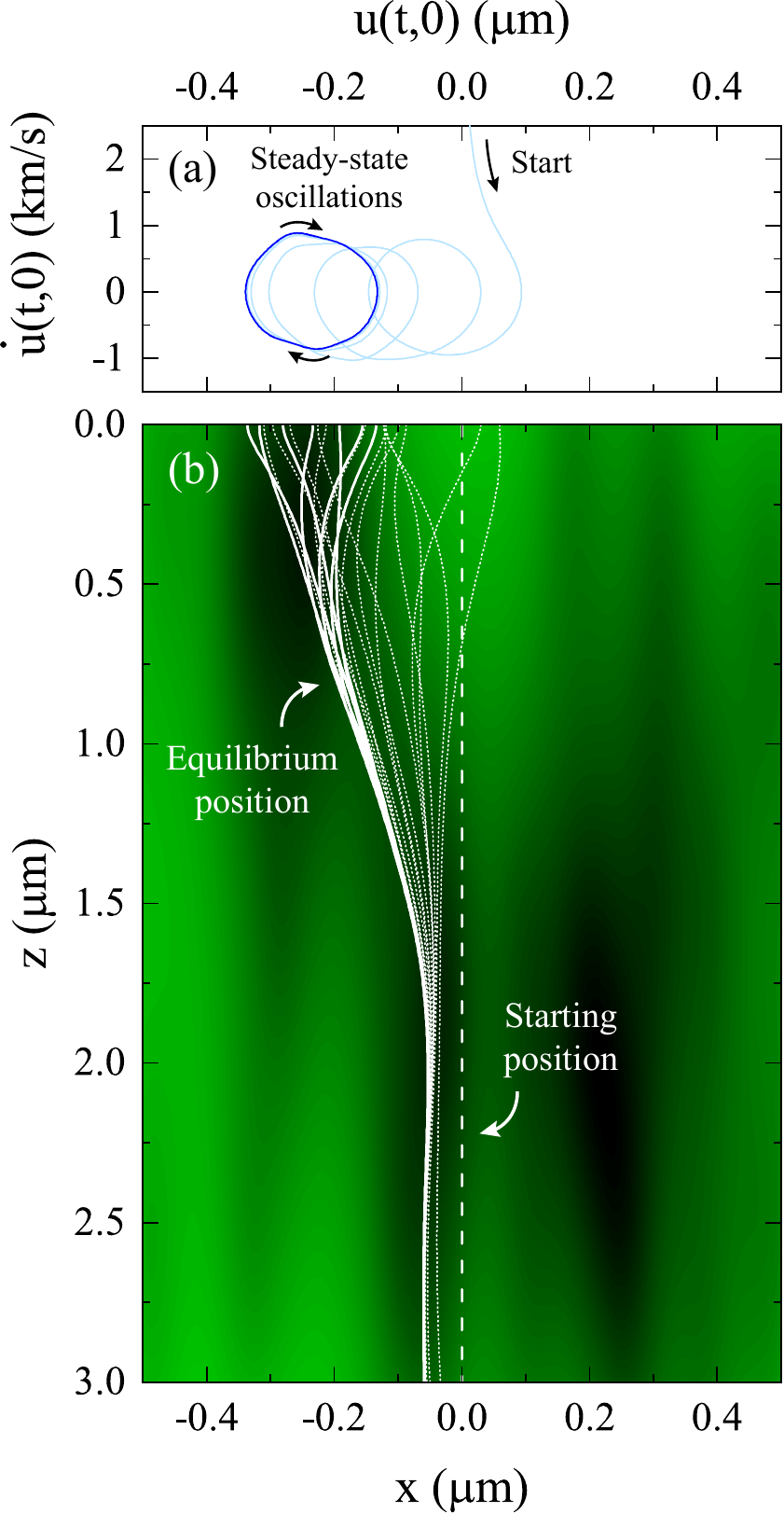}
\caption{Example of vortex dynamics simulation. In (a), the phase portrait of the vortex tip is reported. The arrows indicate the direction of the vortex motion. In light blue the transition from initial condition to steady-state is shown. In (b), the time evolution of the vortex displacement with respect its initial position\textemdash dashed line\textemdash is reported. The solid lines show the vortex motion in the steady-state, while the dotted lines the time evolution from initial condition to steady-state.}
\label{fig.steady_motion}
\end{figure}

The vortex motion equation is rewritten to be solved numerically, by defining lengths in units of $\lambda$ and times in unit of $1/f$: $u=\Tilde{u}\lambda$, $z=\Tilde{z}\lambda$, $X_i=\Tilde{X_i}\lambda$, $Z_i=\Tilde{X_i}\lambda$, and $t=\Tilde{t}/f$. The motion equation then becomes:
\begin{equation}
    \begin{split}
        \eta\lambda f\Dot{\Tilde{u}}\left( \Tilde{t},\Tilde{z}\right)&=\dfrac{\epsilon}{\lambda}\Tilde{u}''\left( \Tilde{t},\Tilde{z}\right)+f_p\left( \Tilde{t},\Tilde{z}\right)+\\
        &+\gamma B_{p}\text{cos}(2\pi\Tilde{t})e^{-\Tilde{z}}\text{.}
    \end{split}
    \label{eq.motion_less}
\end{equation}

The motion equation in Eq.~\ref{eq.motion_less} was solved numerically by means of the method of lines. The spatial derivatives along $z$ are discretized into a grid of $M$ points with a finite elements method, while the time variable is kept continuous. Equation~\ref{eq.motion_less} was thus approximated by a system of $M$ differential equations each describing the time evolution of $\Tilde{u}$ evaluated on one of the $M$ grid points. To solve Eq.~\ref{eq.motion_less}, $M$ was chosen in between 1000 and 10000 depending on the complexity of the pinning landscape.

The calculation is let evolve for several periods of oscillation, until the solution converges to a steady-state condition, identified by a close trajectory in the vortex motion phase portrait, $\Dot{\Tilde{u}}$ versus $\Tilde{u}$.

In Fig.~\ref{fig.steady_motion}(a), an example of the vortex motion phase portrait at the surface ($z=0$) calculated for surface peak field $B_p=50$~mT, $f_{p,ave}=50$~$\mu$N/m, and $l=300$~nm is shown. The motion starts with the vortex in the initial condition ($\Tilde{u}(0,\Tilde{z})=0$) and evolves with time until the steady-state condition is reached.

In Fig.~\ref{fig.steady_motion}(b), the time evolution of the vortex displacement within the pinning landscape is shown. The vortex initiates its motion from the initial condition represented by the straight dashed line at $x=0$. Then, it starts oscillating at the surface while moving towards negative $x$ values attracted by a nearby pinning center (darker region). Because of this, the initial motion of the vortex is chaotic\textemdash as shown by the phase space in Fig.~\ref{fig.steady_motion}(a) and by the dotted lines in Fig.~\ref{fig.steady_motion}(b)\textemdash and eventually converges to steady-state oscillations when the vortex find its equilibrium position attracted by a pinning center, as shown by the solid lines in Fig.~\ref{fig.steady_motion}(b) on the left side of the solution domain.

As briefly described above, the self-stabilization of the vortex to equilibrium according to the pinning landscape is allowed by the boundary condition chosen at the far edge of the solution domain. The condition $\Tilde{u}'(\Tilde{t},\Tilde{z}_{max})=0$ allows the vortex to move and reach an equilibrium position without being constrained to a specific location. 

Simulations were also performed preparing the initial condition by letting the vortex relaxing to equilibrium before applying the driving force. The employment of such simulation strategy resulted in an equivalent outcome as if $\Tilde{u}(0,\Tilde{z})=0$ was used, but required longer time to converge to a steady-state solution. Therefore, all the simulations here presented were performed using the initial condition reported in Eq.~\ref{eq.BC}.

\section{Vortex Surface Resistance}
\begin{figure*}[t]
\centering
\includegraphics[width=17cm]{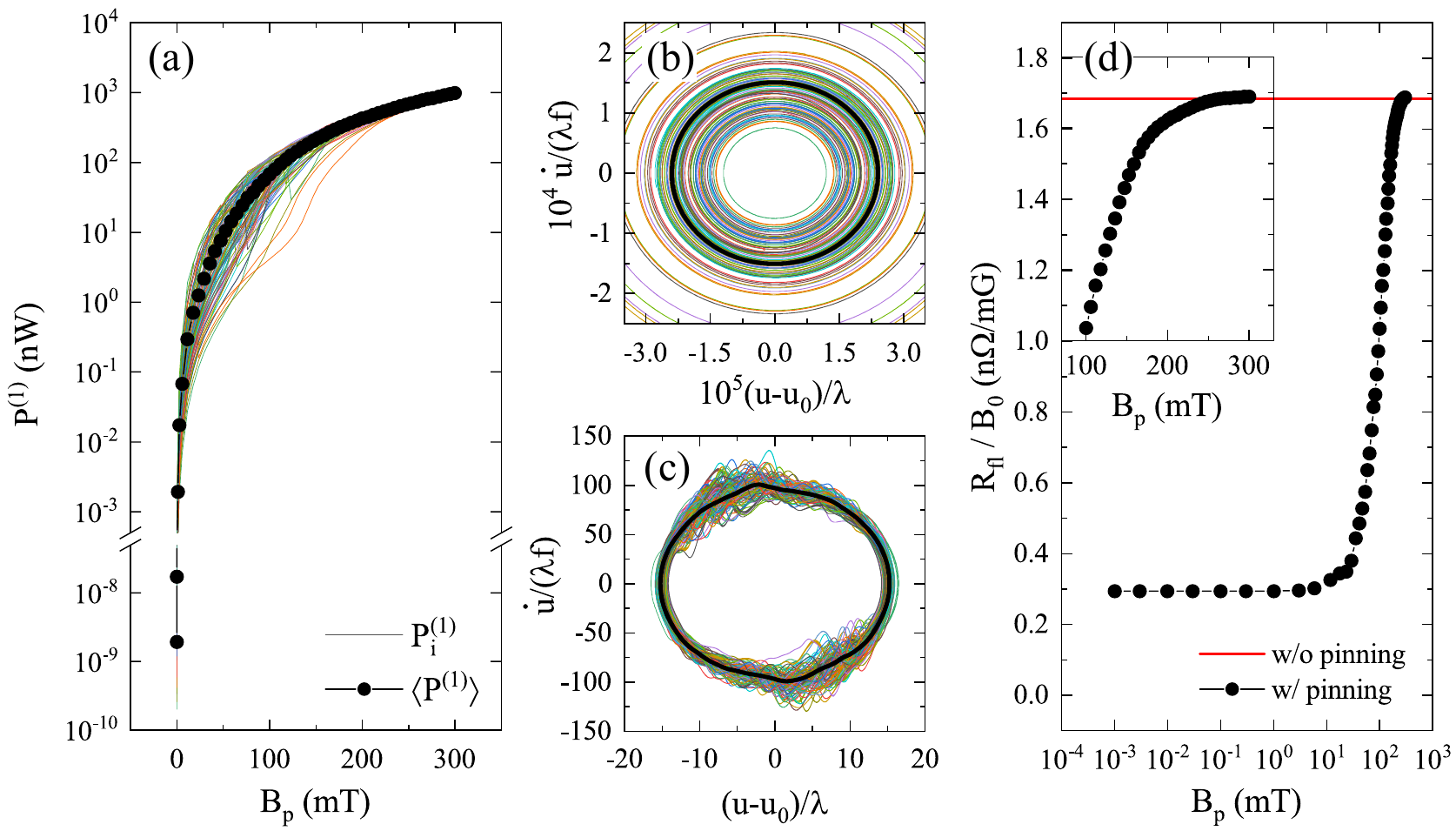}
\caption{Example of vortex surface resistance simulation calculated for $l=300$~nm and $f_{p,ave}=200$~$\mu$N/m as a function of peak field. In graph (a), the average single-vortex power dissipation ($\langle P^{(1)}\rangle$) calculated as a function of the peak magnetic field is reported. In (b) and (c), the average phase portraits calculated over 100 vortex simulations at 0.001~mT and 300~mT respectively is reported. The parameter $u_0$ represent the $x$ offset of the oscillation center. Thin colored solid line represents the individual vortex phase portrait, while the thick solid black line their average. In (d), the simulated $R_{fl}/B_0$ curve is shown with abscissa in semi-logarithmic scale. In the inset, the same curve in the range $1\text{~n}\Omega\text{/mG}\leq R_{fl}/B_0\leq1.8\text{~n}\Omega\text{/mG}$ is shown with abscissa in linear scale.}
\label{fig.simulation}
\end{figure*}
During a FC experiment, we assume that all the magnetic flux applied during cooldown gets trapped in the material and divided into $N_v$ vortexes. Once an electromagnetic field is applied to the material, inducing currents parallel to the surface, vortexes oscillates absorbing part of the electromagnetic energy dissipating power.

The power per unit of surface dissipated by vortex motion is defined as:
\begin{equation}
    \dfrac{dP}{d\Sigma}=\dfrac{1}{2\mu_0^2}R_{fl}B_p^2\text{.}
    \label{eq.powerArea}
\end{equation}

Considering a situation in which the magnetic field $B_p$ and $R_{fl}$ are constant over the surface, $R_{fl}$ can be defined as:
\begin{equation}
    R_{fl}=\dfrac{2\mu_0^2}{B_p^2\Sigma}P\text{,}
    \label{eq.resistance1}
\end{equation}
where $P$ represents the total power dissipated by the collection of the $N_v$ oscillating vortexes.
\begin{figure*}[t]
\centering
\includegraphics[width=17cm]{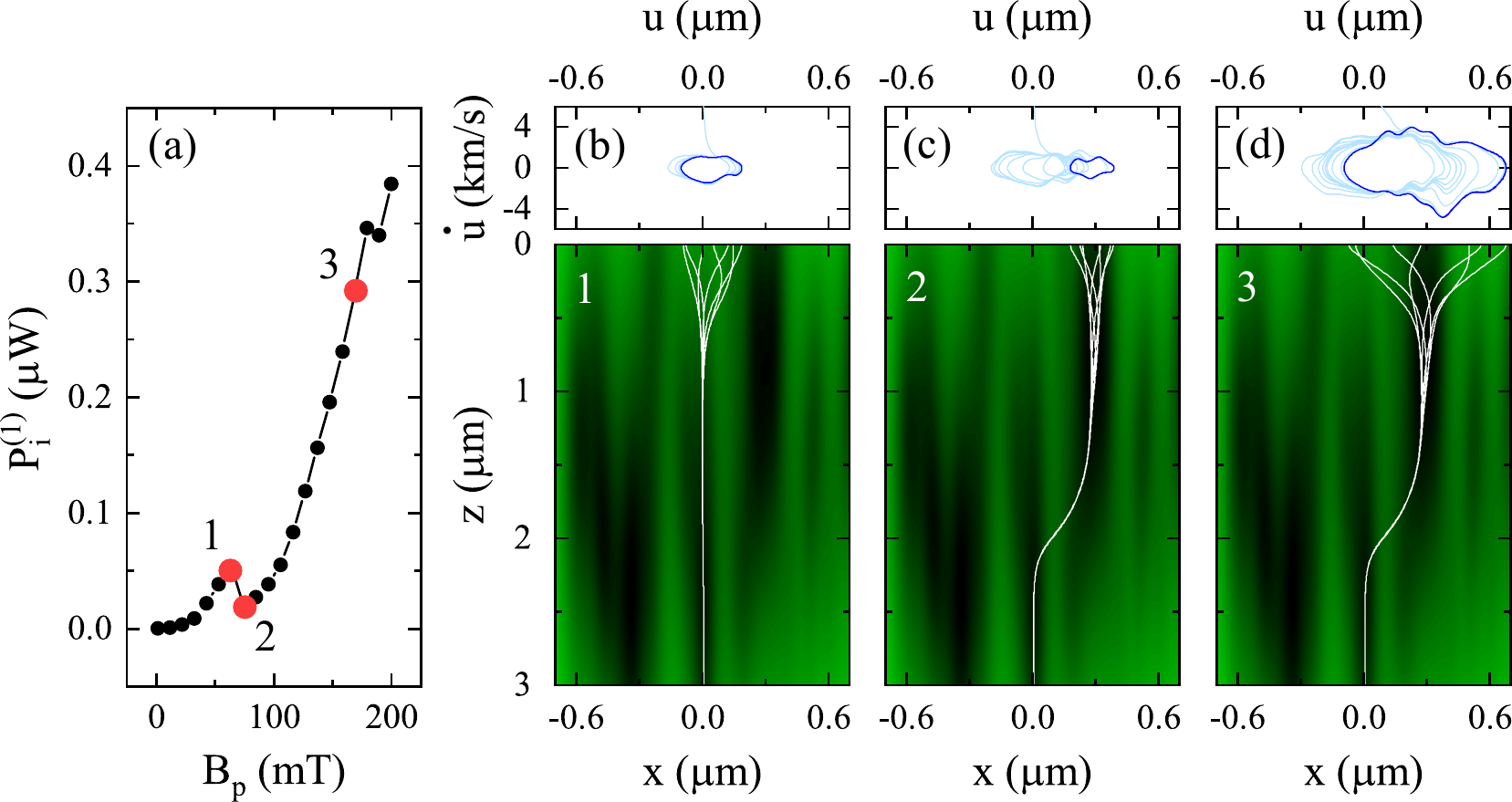}
\caption{Example of single-vortex dynamics as a function of the microwave field. In (a), the single-vortex power dissipation as a function of the peak magnetic field is shown. In (b), (c), and (d), phase portraits and vortex displacements as a function of time calculated at the field level identified by red points and labels 1, 2, and 3 respectively in graph (a) are shown.}
\label{fig.depinning}
\end{figure*}

Because of the random nature of the pinning landscape in real materials, we assume that each vortex oscillates within its local pinning landscape, which differs from vortex to vortex except for its average pinning force. Under this assumption, the power $P$ represents the sum of the single-vortex dissipated powers $P_i^{(1)}$ calculated over the total number of $N_v$ vortexes:
\begin{equation}
\begin{split}
    P&=\sum_i^{N_v}P_i^{(1)}\\
    P_i^{(1)}&=\dfrac{\gamma B_p}{T}\int_0^T \int_0^{z_{max}}\dot{u}_i(t,z)\text{cos}(\omega t)e^{-z/\lambda}dz dt\text{,}
\end{split}
\end{equation}
with $\dot{u}_i(t,z)$ the $i$-th vortex velocity and $T$ the microwave field oscillation period.

The number of vortexes trapped in the material is estimated by assuming that each vortex carries a single flux quantum $\phi_0$ and that vortexes are equally distributed over the area $\Sigma$. In turn, the applied magnetic flux during the cooldown $B_0 \Sigma$ is divided into $N_v=B_0\Sigma/\phi_0$ vortexes. Thus Eq.~\ref{eq.resistance1} can be rewritten as:
\begin{equation}
\begin{split}
        R_{fl}&=\dfrac{2\mu_0^2B_0}{\phi_0 B_p^2}\dfrac{\sum_i^{N_v}P_i^{(1)}}{N_v}\\
        &=\dfrac{2\mu_0^2B_0}{\phi_0 B_p^2}\langle P^{(1)} \rangle\text{,}
\end{split}
\end{equation}
where $\langle P^{(1)} \rangle$ represents the average of the $N_v$ vortexes dissipated power.

By applying the same definitions used in Eq.~\ref{eq.motion_less}, we can rewrite the single-vortex average dissipated power as:
\begin{equation}
\begin{split}
   \langle P^{(1)}\rangle&=\dfrac{\phi_0 B_p\lambda f}{\mu_0 N_v}\times\\
   &\times\sum_i^{N_v}\int_0^1\int_0^{\Tilde{z}_{max}} \dot{\Tilde{u}}_i\left(\Tilde{t},\Tilde{z}\right)\text{cos}\left(2\pi\Tilde{t}\right)e^{-\Tilde{z}}d\Tilde{z} d\Tilde{t}\text{.}    
\end{split}
\label{eq.resistance2}
\end{equation}

A Montecarlo approach was implemented to estimate $\langle P^{(1)}\rangle$ by considering a statistical sample of 100 vortexes. In the range of field probed\textemdash tens to hundreds of mG\textemdash and areas considered\textemdash order of the square meter, vortexes exceeds the hundred million unities, making Eq.~\ref{eq.resistance2} impossible to compute by means of a desktop computer if the total number of vortexes involved was considered.

The vortex motion equation in Eq.~\ref{eq.motion_less} was computed for each of the 100 vortexes generating each time an unique random pinning landscape. $P^{(1)}_i$ was calculated once the solution to Eq.~\ref{eq.motion_less} approached the steady-state. The estimation of $\langle P^{(1)}\rangle$ was carried out by averaging the outcome of the single-vortex simulations. An example of calculation for $l=300$~nm and $f_{p,ave}=200$~$\mu$N/m is reported in Fig.~\ref{fig.simulation}(a).

The colored solid lines represents the single-vortex power $P^{(1)}_i$ as a function of $B_p$ calculated each time for a random pinning landscape, for a total of 100 vortexes. The solid line with black dots represents instead the average single-vortex dissipated power $\langle P^{(1)}\rangle$.

In Fig.~\ref{fig.simulation}(b) and (c) the average phase portrait of vortex oscillation calculated over 100 vortexes, respectively at $B_p=0.001$~mT and 300~mT, is shown by the thick black solid line. The thin coloured lines represent the individual single-vortex phase portraits. Fig.~\ref{fig.simulation}(d) and its inset show instead the calculated normalized vortex surface resistance with abscissa in semi-logarithmic and linear scales, respectively.

At very low microwave fields, the vortex performs oscillations with small amplitude, since well constrained inside a local pinning potential. In this regime, the pinning potential can be approximated by a parabolic function, and the vortex motion equation (Eq.~\ref{eq.motion_less}) becomes linear. Thus, the calculated vortex response is linear, implying an elliptical phase portrait and constant surface resistance (Fig.~\ref{fig.simulation}(b) and (d)).

At high microwave fields, the vortex surface resistance tends to saturate to a constant value equal to the surface resistance in absence of pinning. In this case, even if each vortex shows a non-linear response, the average phase portrait tends to an ellipse (Fig.~\ref{fig.simulation}(c)), meaning that the global response becomes linear for higher microwave fields. Pinning is overcame when vortexes are driven at higher microwave fields and the surface resistance saturates to a constant value equal to the vortex surface resistance in absence of pinning, as shown in the inset of Fig.~\ref{fig.simulation}(d).

Such saturation behavior of the vortex surface resistance above a certain microwave field level was also predicted as generated by viscous losses in Ref.~\onlinecite{Liarte_PRApplied_2018}.
\begin{figure*}[t]
\centering
\includegraphics[width=17cm]{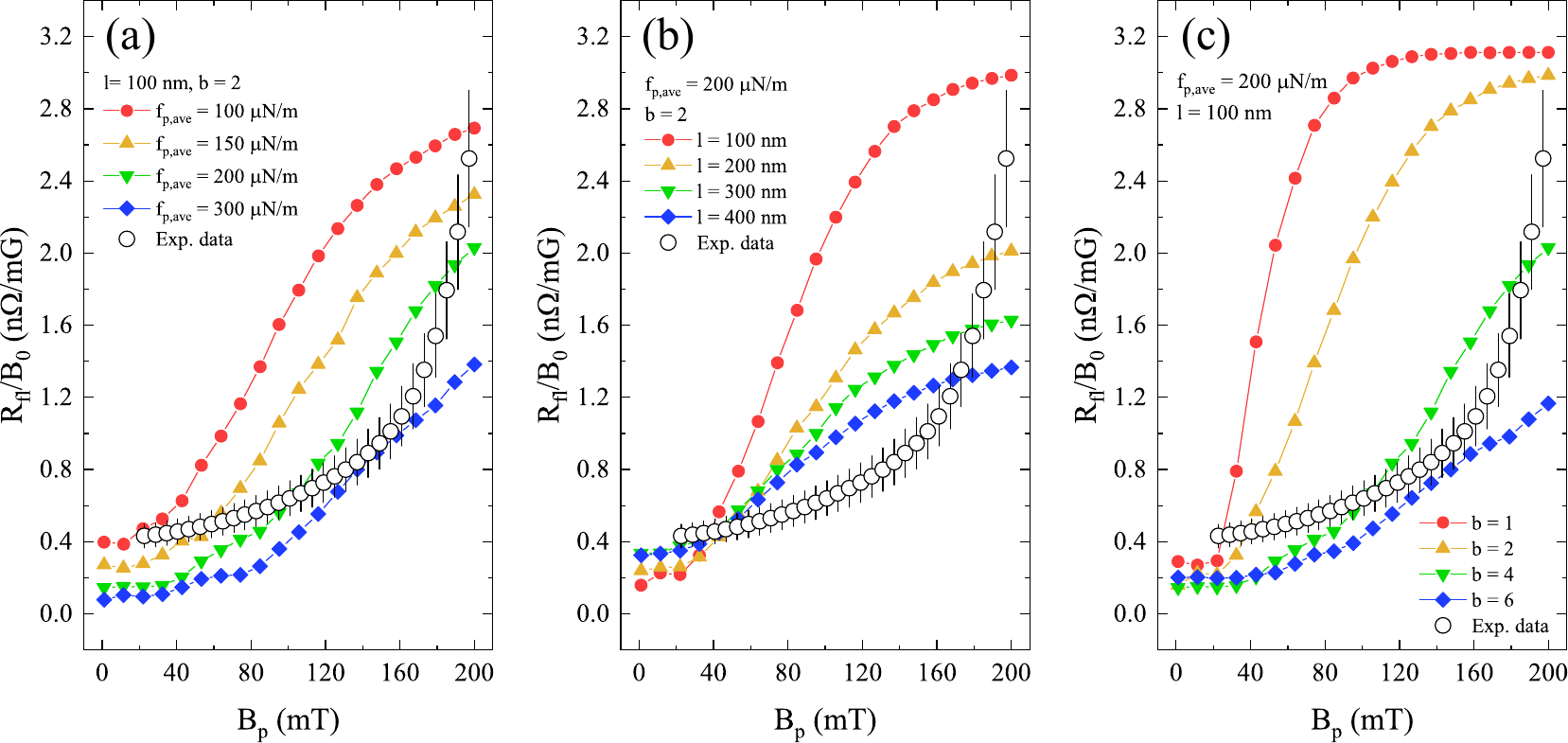}
\caption{Comparison of simulated normalized vortex surface resistance with experimental data. In (a), simulations for different average pinning force at fixed mean-free-path and fixed factor $b$. In (b), simulations as a function of the mean-free-path value for fixed average pinning force and factor $b$. In (c), simulations as a function of the factor $b$ for fixed average pinning force and mean-free-path.}
\label{fig.simulations}
\end{figure*}

In summary, for low microwave fields pinning can be treated as a parabolic potential\textemdash the pinning force is linear and proportional to the vortex displacement, while for high microwave fields pinning becomes negligible. In both cases, the vortex dynamics is approximated to be linear leading to a constant surface resistance.

\section{Microwave Depinning\label{sub.jump}}
Simulations of single-vortex dynamics allow to shed light into the mechanism behind the dependence of $R_{fl}$ against the amplitude of the microwave drive.

In Fig.~\ref{fig.depinning}, an example of single-vortex dynamics is reported. The single-vortex power dissipation ($P^{(1)}_i$) curve shown in Fig.~\ref{fig.depinning}(a) was selected from Fig.~\ref{fig.simulation}(a). The curve presents a step like behavior (see highlighted points 1 and 2) as a function of $B_p$. Such effect is consequence of a ``vortex jump'', where the vortex gets depinned from a location and captured by a stronger pinning center nearby, as a consequence of microwave depinning.

Simulations in Fig.~\ref{fig.depinning}(b), (c), and (d) show the vortex jump behavior just discussed and the microwave depinning phenomenon. The darker blue line in the phase portraits represents steady-state oscillations, while the simulations underneath show the time evolution of the vortex oscillation in the steady-state only (for clarity). The numbers 1, 2, and 3 indicate the $B_p$ value at which the simulations were done (63~mT, 75~mT, and 170~mT respectively) and highlight the respective $P_i^{(1)}$ values in Fig.~\ref{fig.depinning}(a).

At $B_p=63$~mT (point 1), the vortex tip oscillates almost depinned from the local pinning center. As soon as the $B_p$ is increased to $75$~mT (point 2), the oscillation widens and the vortex starts to interact with a nearby pinning center, which eventually captures it.

Comparing the two simulations (point 1 and 2), appears clear that because of the jump from one pinning center to a stronger one, the amplitude of the steady-state oscillations are reduced and so it is the oscillation velocity. As a result, the power dissipation is reduced accordingly, generating this abrupt step characteristic in the $B_p$ dependence of $P^{(1)}_i$.

The progressive microwave depinning phenomenon is appreciable by comparing the simulations of point 2 and 3. At $B_p=75$~mT (point 2) the vortex oscillation is well constrained by the pinning center, while for higher peak magnetic field values (e.g 170~mT, point 3) the oscillation widens substantially reaching areas outside the local pinning potential. If the conditions are right, vortex jump induced by microwave depinning can repeat at higher fields, as shown in Fig.~\ref{fig.depinning}(a) around 180~mT.

Each vortex might experience jumps and depinning at different $B_p$ values depending on the local pinning landscape, producing an unique $P_i^{(1)}$ versus $B_p$ characteristic. In the global picture, the microwave amplitude dependence of $R_{fl}$ is defined by the collective vortex motion and the abrupt steps expected in the $P_i^{(1)}$ due to individual vortex jumps are averaged out.

\section{Comparison with Experimental Data}
\subsection{Microwave Data}
In Fig.~\ref{fig.simulations}(a), (b), and (c), the trapped flux surface resistance measured experimentally is compared to several simulations of trapped flux surface resistance calculated for increasing values of average pinning force $f_{p,ave}$, mean-free-path $l$, and pinning center size $b\xi$, respectively.

As shown in Fig.~\ref{fig.simulations}(a), for larger average pinning force, the $B_p$ value at which saturation is reached increases and the overall $R_{fl}$ curve is shifted to lower values. The effect of different mean-free-path is instead shown in Fig.~\ref{fig.simulations}(b). The flux-flow viscosity decreases for low $l$ values, the vortex moves more freely and the saturation value increases. The slope of the $R_{fl}$ microwave field amplitude dependence is instead dictated by the dimension of the pinning centers, as shown in Fig.~\ref{fig.simulations}(c). The smaller the pinning center, the steeper the trend and the earlier the saturation is reached.

Since the pinning landscape is unknown a-priori, it is virtually impossible to generate quantitative simulations of vortex surface resistance as a function of the microwave field amplitude. Nevertheless, the simulations can qualitatively describe the experimental data. Based on the simulations results, progressive microwave depinning is the most likely culprit for the microwave field dependence of the vortex surface resistance, the higher the driving force, the less constrained the vortexes and wider the oscillations, leading to higher resistance. 

orHowever, the experimental data does not present saturation at high $B_p$ values, contradicting the simulations. This might occur because: i) saturation is reached at higher fields, and ii) others non-linearity play a role in the vortex dynamics. The first possibility is related to the fact that pinning landscape and mean-free-path used in the simulations may not be accurate enough\textemdash pinning landscape is unknown a-priori and cavities treated as the one under study are characterized by $l$ varying with depth, with much shorter values at the surface.\cite{Romanenko_APL_2014} The second possibility is instead substantiated by the fact that in specific conditions the vortex flux-flow viscosity decreases with the vortex velocity, introducing another non-linearity in the motion equation. Such effect, first described by Larkin and Ovchinnikov,\cite{Larkin_JETP_1976} will be discussed in Section~\ref{sec.LO}.

\subsection{Geometrical Considerations}
Due to the not planar geometry of the SRF cavity, the vortex surface resistance is not constant across the surface, it instead varies as described previously in Section~\ref{sec.data}.

The temperature variation $\Delta T_{out}$\textemdash calculated as the FC and ZFC Tmap data subtraction\textemdash is directly related to the surface resistance due to vortexes. The surface resistance is directly proportional to the power dissipated, as shown in Eq.~\ref{eq.resistance1}, which, in turn, is directly proportional to the temperature variation on the outside surface of the cavity through the thermal transport law, i.e. $R_{fl}\propto \Delta T_{out}/B_p^2$.

\begin{figure}[t]
\centering
\includegraphics[width=8.5cm]{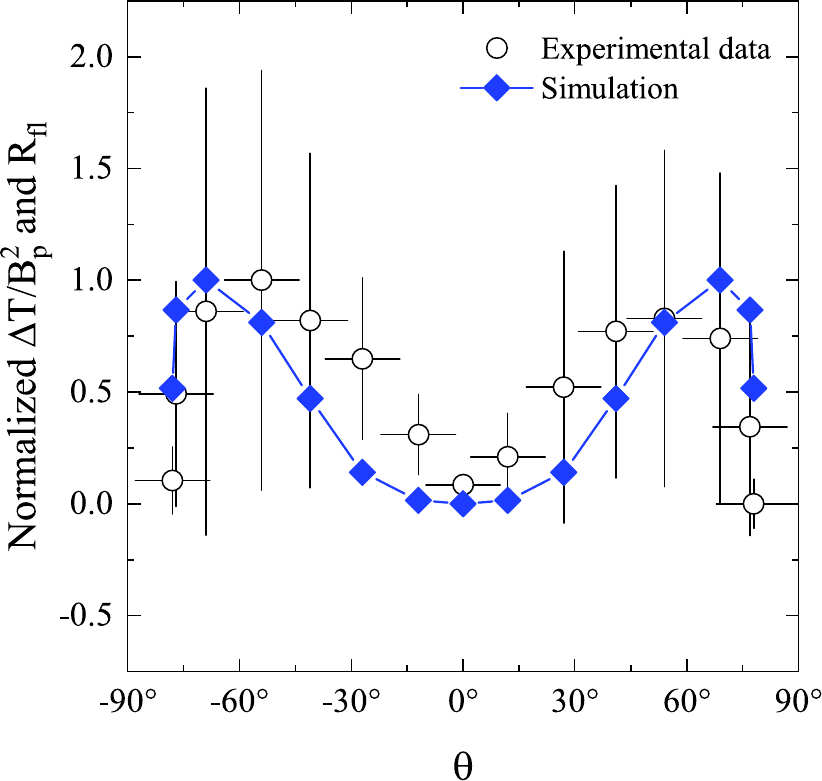}
\caption{Normalized experimental data $\Delta T/B_p^2$ and normalized simulated $R_{fl}$ plotted against the angle between microwave currents and magnetic flux directions.}
\label{fig.angle}
\end{figure}

Figure~\ref{fig.angle} reports the normalized data shown in Fig.~\ref{fig.Tmap}(c) averaged over the board number as a function of the angle $\theta$ between $B_0\hat{z}$ and the microwave currents flowing parallel to the cavity surface. $\theta$ was estimated as the angle between $B_0\hat{z}$ and the cavity surface tangent at each thermometer location.

The error bars are defined to be the standard deviation of the $\Delta T_{out}/B_p^2$ average value in ordinate, while for the values in abscissa the error was estimated a-posteriori to be $10\degree$.

The experimental data is compared against the model here presented, where the Lorentz force $\text{sin}(\theta)$ dependence is taken into consideration, as well as the local surface current distribution at the cavity surface reported in Fig.~\ref{fig.Tmap}(b).

The calculation was performed with the same Montecarlo approach described above, where every point is calculated averaging the dissipated power over 100 vortexes. The simulation was carried out assuming $l=100$~nm, $f_{p,ave}=200$~$\mu$N/m, and $b=4$, but any other set of parameters returned comparable results. In order to compare experimental data and simulation, all the quantities in ordinate are normalized between 0 and 1.

The angle dependence calculated is in good agreement to the experimental data. At $\theta=0$\degree, the surface resistance is zero since the trapped field is parallel to the microwave currents, while at $\theta\simeq80$\degree, the surface resistance decreases because the microwave current drops rapidly to zero approaching the cavity irises.

\section{Larkin-Ovchinnikov Instability\label{sec.LO}}
Larkin-Ovchinnikov (LO) instability\cite{Larkin_JETP_1976} is a phenomenon of non-equilibrium related to vortex motion. Inside the vortex core superconductivity is suppressed and quasi-particles can occupy virtually any energy level. During vortex motion the energy of quasi-particles increases and when reaches $\sim\Delta$, quasi-particles diffuse out of the vortex into the superconductor, the vortex core gets depleted, shrinks in dimension, and the flux-flow viscosity decreases. 

In the LO theory framework, the critical velocity is defined as\cite{Larkin_JETP_1976,Grimaldi_PRB_2010,Leo_PRB_2011}
\begin{equation}
    v^*\propto\tau_\varepsilon^{-1/2}\left(1-\dfrac{T}{T_c}\right)^{1/4}\text{,}
    \label{eq.v}
\end{equation}
where $\tau_\varepsilon^{-1}$ is the rate at which quasi-particles relax back into the vortex core. The shorter $\tau_\varepsilon$, the faster the vortex must travel to meet the instability onset.

Experimental studies on Nb thin films\cite{Grimaldi_PRB_2010} showed that the LO instability critical velocity $v^*$ is function of the magnetic field trapped in the sample. The low $B_0$ region ($B_0\leq100$~mG) which is of our interest is still poorly explored. However, it was shown that, in Nb, $v^*$ increases for very low fields,\cite{Grimaldi_PRB_2010, Dobrovolskiy_SUST_2017} reaching values of the order of couple of km/h for $B_0\sim1$~G.

The theory developed by Larkin and Ovchinnikov is strictly applicable to $T\simeq T_c$. However, the underlying decreasing of the flux-flow viscosity as a function of the vortex velocity is a feature observed also at low temperatures,\cite{Leo_PRB_2011} and therefore relevant in Nb SRF cavities operating at 2~K, especially at high gradients. In order to expand the applicability of such formulation to low temperatures, the quasi-particle relaxation rate $\tau_\varepsilon^{-1}$ shall be defined as the sum $\tau_s^{-1}+\tau_r^{-1}$, respectively scattering and recombination rates.\cite{Leo_PRB_2011}

Generally speaking, at low temperatures ($T<T_c/2$), $\tau_s^{-1}$ and $\tau_r^{-1}$ decrease due to the lower number of phonons and quasi-particles, respectively.\cite{Kaplan_PRB_1976} However, if $\tau_s^{-1}$ is governed by scattering with impurities or lattice defects, it becomes temperature-independent. The occurrence of LO instability should then be more relevant in clean niobium, where the quasi-particle relaxation time is not limited by scattering with impurities and results longer.

In the present study, we discuss trapped flux surface resistance in cavities with low mean-free-path, prepared with a mild baking treatment (Section~\ref{sec.data}), that was proven to increase the concentration of interstitial oxygen and vacancies in the material.\cite{Ciovati_JAppPhys_2004,Romanenko_APL_2013,Romanenko_SRF_2019} We can then assume that the relaxation rate is most likely dominated by scattering with impurities, and therefore that $v^*$ increases with $T$ as expected by Eq.~\ref{eq.v}.

Hence, if occurring, LO instability would be appreciable at higher microwave fields, introducing deviations from the expected plateau in the surface resistance predicted by this work. However, since no saturation in the experimental data was observed at high microwave field amplitudes, LO instability cannot be totally ruled out, inasmuch saturation might be masked by such phenomenon. 

\section{Conclusions}
In this work, the vortex surface resistance as a function of the microwave field amplitude is simulated by means of a Montecarlo approach aimed to solve the motion equation for a collection of vortexes each oscillating within a randomly defined pinning landscape.

Simulations suggest that at low microwave amplitudes vortexes are well constrained within a local pinning potential and their response is linear, leading to a constant surface resistance. For increasing field amplitudes vortexes may oscillate out of their local pinning potential and possibly jump from their local to a neighbouring pinning center introducing non-linearity. The effect of such phenomena leads to an increase of the vortex surface resistance until the average response approaches a linear behavior for higher field amplitudes, where the average vortex motion becomes insensitive to the pinning landscape. When this happens the surface resistance saturates to a constant value.

Slope, saturation value, and saturation onset are closely related to the pinning landscape\textemdash not known a-priori\textemdash within which vortexes oscillate. Quantitative description of $R_{fl}$ as a function of $B_p$ is then virtually impossible. Nevertheless, the experimental data are in qualitative agreement with the simulations especially the data as a function of the angle between microwave current and vortex magnetic flux direction.

The experimental data as a function of the microwave field amplitude do not show saturation as instead expected from simulations. As described above, this might be due to the not enough accurate choice of pinning parameters and mean-free-path value in the simulations, or to the onset of LO instability at high fields.

Concluding, this work allowed to gain insights into vortex dynamics at high amplitude microwave drive and on the transition from low to high field amplitudes. The microscopic nature the microwave field dependence of the vortex surface resistance was presented and attributed to microwave depinning of vortexes. Appreciable contributions from thermal feedback effects for the magnetic field values and microwave amplitudes explored experimentally\textemdash $B_0<100$~mG and $B_p<200$~mT\textemdash were instead ruled out.

\section{Acknowledgements}
We thank Alex Romanenko and Martina Martinello for the insightful discussions that help the development of this work. This work was supported by the United States Department of Energy, Office of High Energy Physics. Fermilab is operated by Fermi Research Alliance, LLC under Contract No. DE-AC02-07CH11359 with the United States Department of Energy.

\appendix
\section{Penetration Depth and Coherence Length Calculation}
Since the numeric calculation is performed for finite vales of electron mean-free-path ($l$) value, the penetration depth of niobium was calculated by means of the spatial Fourier transform of the Pippard's Kernel~\cite{Pippard_ProcRSocLond_1953} $K(q,l,T)$, as described by Tinkham,~\cite{Tinkham_Book} that closely approximates the BCS~\cite{Bardeen_PR_1957} results.

For isotropic superconductors, such as niobium, $K(q,l,T)$ is defined as:
\begin{equation}
\begin{split}
    &K(q,l,T)=\dfrac{\xi(l,T)}{\xi_0(T)\lambda_L(T)^2}\bigg\{ 
\dfrac{3}{2(q\xi(l,T))^3}+\\
&+\left[(1+q^2\xi(l,T)^2)\text{tan}^{-1}q\xi(l,T)-q\xi(l,T)\right]\bigg\}\text{,}
\end{split}
\end{equation}
where $\lambda_L(T)$ is approximated by the two-fluid model definition $\lambda_L(T)=\lambda_L(0)\left[1-(T/T_c)^4\right]^{-1/2}$,~\cite{Gorter_Physica_1934} and the $T$ dependence of $\xi$ and $\xi_0$ is derived from the BCS theory:~\cite{Bardeen_PR_1957}
\begin{equation}
\begin{split}
    \xi(l,T)&=\left[\dfrac{J(0,T)}{\xi_0(0)}+\dfrac{1}{l}\right]^{-1}\\
    \xi_0(T)&=\dfrac{\xi_0(0)}{J(0,T)}\text{.}
\end{split}
\end{equation}

Here, $J(0,T)$ is the real-space Kernel as defined as:~\cite{Bardeen_PR_1957,Suter_PRB_2005}
\begin{equation}
    J(0,T)=\left[\dfrac{\lambda_L(T)}{\lambda_L(0)}\right]^2\dfrac{\Delta(T)}{\Delta(0)}\text{tanh}\dfrac{\Delta(T)}{2\kappa_BT}\text{,}
\end{equation}
where $\Delta(T)/\Delta(0)=\left[\text{cos}\left( \pi/2\left( T/T_c\right)^2\right)\right]^{1/2}$,\cite{Sheehan_PR_1966} and $\Delta(0)$ the energy gap at $T=0$~K.

The penetration depth $\lambda(T)$ is then calculated numerically for specular reflection at the boundary as:
\begin{equation}
    \lambda(l,T)=\dfrac{2}{\pi}\int_{0}^{\infty}\dfrac{dq}{K(q,l,T)+q^2}\text{.}
\end{equation}

The parameters $\lambda_L(0)$ and $\xi_0(0)$ represent the London penetration depth and the coherence length in the clean limit for $T=0$, respectively 39~nm and 38~nm.\cite{Maxfield_PhysRev_1965}

\bibliography{Bibliography}

\begin{thebibliography}{67}%
\makeatletter
\providecommand \@ifxundefined [1]{%
 \@ifx{#1\undefined}
}%
\providecommand \@ifnum [1]{%
 \ifnum #1\expandafter \@firstoftwo
 \else \expandafter \@secondoftwo
 \fi
}%
\providecommand \@ifx [1]{%
 \ifx #1\expandafter \@firstoftwo
 \else \expandafter \@secondoftwo
 \fi
}%
\providecommand \natexlab [1]{#1}%
\providecommand \enquote  [1]{``#1''}%
\providecommand \bibnamefont  [1]{#1}%
\providecommand \bibfnamefont [1]{#1}%
\providecommand \citenamefont [1]{#1}%
\providecommand \href@noop [0]{\@secondoftwo}%
\providecommand \href [0]{\begingroup \@sanitize@url \@href}%
\providecommand \@href[1]{\@@startlink{#1}\@@href}%
\providecommand \@@href[1]{\endgroup#1\@@endlink}%
\providecommand \@sanitize@url [0]{\catcode `\\12\catcode `\$12\catcode
  `\&12\catcode `\#12\catcode `\^12\catcode `\_12\catcode `\%12\relax}%
\providecommand \@@startlink[1]{}%
\providecommand \@@endlink[0]{}%
\providecommand \url  [0]{\begingroup\@sanitize@url \@url }%
\providecommand \@url [1]{\endgroup\@href {#1}{\urlprefix }}%
\providecommand \urlprefix  [0]{URL }%
\providecommand \Eprint [0]{\href }%
\providecommand \doibase [0]{http://dx.doi.org/}%
\providecommand \selectlanguage [0]{\@gobble}%
\providecommand \bibinfo  [0]{\@secondoftwo}%
\providecommand \bibfield  [0]{\@secondoftwo}%
\providecommand \translation [1]{[#1]}%
\providecommand \BibitemOpen [0]{}%
\providecommand \bibitemStop [0]{}%
\providecommand \bibitemNoStop [0]{.\EOS\space}%
\providecommand \EOS [0]{\spacefactor3000\relax}%
\providecommand \BibitemShut  [1]{\csname bibitem#1\endcsname}%
\let\auto@bib@innerbib\@empty
\bibitem [{\citenamefont {Abrikosov}(1957)}]{Abrikosov_ZETF_1957}%
  \BibitemOpen
  \bibfield  {author} {\bibinfo {author} {\bibfnamefont {A.~A.}\ \bibnamefont
  {Abrikosov}},\ }\href@noop {} {\bibfield  {journal} {\bibinfo  {journal} {Zh.
  Eksp. Teor. Fiz.}\ }\textbf {\bibinfo {volume} {32}},\ \bibinfo {pages}
  {1442} (\bibinfo {year} {1957})},\ \translation{A. A. Abrikosov, Soviet
  Phys.\textemdash JETP \textbf{5}, 1174 (1957)}\BibitemShut {NoStop}%
\bibitem [{\citenamefont {Shubnikov}\ \emph {et~al.}(1937)\citenamefont
  {Shubnikov}, \citenamefont {Khotkevich}, \citenamefont {Shepelev},\ and\
  \citenamefont {Riabinin}}]{Shubnikov_ZETF_1937}%
  \BibitemOpen
  \bibfield  {author} {\bibinfo {author} {\bibfnamefont {L.~V.}\ \bibnamefont
  {Shubnikov}}, \bibinfo {author} {\bibfnamefont {V.~I.}\ \bibnamefont
  {Khotkevich}}, \bibinfo {author} {\bibfnamefont {Y.~D.}\ \bibnamefont
  {Shepelev}}, \ and\ \bibinfo {author} {\bibfnamefont {Y.~N.}\ \bibnamefont
  {Riabinin}},\ }\href@noop {} {\bibfield  {journal} {\bibinfo  {journal} {Zh.
  Eksp. Teor. Fiz.}\ }\textbf {\bibinfo {volume} {7}},\ \bibinfo {pages} {221}
  (\bibinfo {year} {1937})}\BibitemShut {NoStop}%
\bibitem [{\citenamefont {Romanenko}\ \emph
  {et~al.}(2014{\natexlab{a}})\citenamefont {Romanenko}, \citenamefont
  {Grassellino}, \citenamefont {Melnychuk},\ and\ \citenamefont
  {Sergatskov}}]{Romanenko_JAP_2014}%
  \BibitemOpen
  \bibfield  {author} {\bibinfo {author} {\bibfnamefont {A.}~\bibnamefont
  {Romanenko}}, \bibinfo {author} {\bibfnamefont {A.}~\bibnamefont
  {Grassellino}}, \bibinfo {author} {\bibfnamefont {O.}~\bibnamefont
  {Melnychuk}}, \ and\ \bibinfo {author} {\bibfnamefont {D.~A.}\ \bibnamefont
  {Sergatskov}},\ }\href@noop {} {\bibfield  {journal} {\bibinfo  {journal} {J.
  Appl. Phys.}\ }\textbf {\bibinfo {volume} {115}},\ \bibinfo {pages} {184903}
  (\bibinfo {year} {2014}{\natexlab{a}})}\BibitemShut {NoStop}%
\bibitem [{\citenamefont {Romanenko}\ \emph
  {et~al.}(2014{\natexlab{b}})\citenamefont {Romanenko}, \citenamefont
  {Grassellino}, \citenamefont {Crawford}, \citenamefont {Sergatskov},\ and\
  \citenamefont {Melnychuk}}]{Romanenko_APL_2014_2}%
  \BibitemOpen
  \bibfield  {author} {\bibinfo {author} {\bibfnamefont {A.}~\bibnamefont
  {Romanenko}}, \bibinfo {author} {\bibfnamefont {A.}~\bibnamefont
  {Grassellino}}, \bibinfo {author} {\bibfnamefont {A.~C.}\ \bibnamefont
  {Crawford}}, \bibinfo {author} {\bibfnamefont {D.~A.}\ \bibnamefont
  {Sergatskov}}, \ and\ \bibinfo {author} {\bibfnamefont {O.}~\bibnamefont
  {Melnychuk}},\ }\href@noop {} {\bibfield  {journal} {\bibinfo  {journal}
  {Appl. Phys. Lett.}\ }\textbf {\bibinfo {volume} {105}},\ \bibinfo {pages}
  {234103} (\bibinfo {year} {2014}{\natexlab{b}})}\BibitemShut {NoStop}%
\bibitem [{\citenamefont {Posen}\ \emph {et~al.}(2016)\citenamefont {Posen},
  \citenamefont {Checchin}, \citenamefont {Crawford}, \citenamefont
  {Grassellino}, \citenamefont {Martinello}, \citenamefont {Melnychuk},
  \citenamefont {Romanenko}, \citenamefont {Sergatskov},\ and\ \citenamefont
  {Trenikhina}}]{Posen_JAP_2016}%
  \BibitemOpen
  \bibfield  {author} {\bibinfo {author} {\bibfnamefont {S.}~\bibnamefont
  {Posen}}, \bibinfo {author} {\bibfnamefont {M.}~\bibnamefont {Checchin}},
  \bibinfo {author} {\bibfnamefont {A.~C.}\ \bibnamefont {Crawford}}, \bibinfo
  {author} {\bibfnamefont {A.}~\bibnamefont {Grassellino}}, \bibinfo {author}
  {\bibfnamefont {M.}~\bibnamefont {Martinello}}, \bibinfo {author}
  {\bibfnamefont {O.~S.}\ \bibnamefont {Melnychuk}}, \bibinfo {author}
  {\bibfnamefont {A.}~\bibnamefont {Romanenko}}, \bibinfo {author}
  {\bibfnamefont {D.~A.}\ \bibnamefont {Sergatskov}}, \ and\ \bibinfo {author}
  {\bibfnamefont {Y.}~\bibnamefont {Trenikhina}},\ }\href@noop {} {\bibfield
  {journal} {\bibinfo  {journal} {J. Appl. Phys}\ }\textbf {\bibinfo {volume}
  {119}},\ \bibinfo {pages} {213903} (\bibinfo {year} {2016})}\BibitemShut
  {NoStop}%
\bibitem [{\citenamefont {Martinello}\ \emph {et~al.}(2016)\citenamefont
  {Martinello}, \citenamefont {Grassellino}, \citenamefont {Checchin},
  \citenamefont {Romanenko}, \citenamefont {Melnychuk}, \citenamefont
  {Sergatskov}, \citenamefont {Posen},\ and\ \citenamefont
  {Zasadzinski}}]{Martinello_APL_2016}%
  \BibitemOpen
  \bibfield  {author} {\bibinfo {author} {\bibfnamefont {M.}~\bibnamefont
  {Martinello}}, \bibinfo {author} {\bibfnamefont {A.}~\bibnamefont
  {Grassellino}}, \bibinfo {author} {\bibfnamefont {M.}~\bibnamefont
  {Checchin}}, \bibinfo {author} {\bibfnamefont {A.}~\bibnamefont {Romanenko}},
  \bibinfo {author} {\bibfnamefont {O.}~\bibnamefont {Melnychuk}}, \bibinfo
  {author} {\bibfnamefont {D.~A.}\ \bibnamefont {Sergatskov}}, \bibinfo
  {author} {\bibfnamefont {S.}~\bibnamefont {Posen}}, \ and\ \bibinfo {author}
  {\bibfnamefont {J.~F.}\ \bibnamefont {Zasadzinski}},\ }\href@noop {}
  {\bibfield  {journal} {\bibinfo  {journal} {Appl. Phys. Lett.}\ }\textbf
  {\bibinfo {volume} {109}},\ \bibinfo {pages} {062601} (\bibinfo {year}
  {2016})}\BibitemShut {NoStop}%
\bibitem [{\citenamefont {Mattis}\ and\ \citenamefont
  {Bardeen}(1958)}]{Mattis_PhysRev_1958}%
  \BibitemOpen
  \bibfield  {author} {\bibinfo {author} {\bibfnamefont {D.~C.}\ \bibnamefont
  {Mattis}}\ and\ \bibinfo {author} {\bibfnamefont {J.}~\bibnamefont
  {Bardeen}},\ }\href@noop {} {\bibfield  {journal} {\bibinfo  {journal} {Phys.
  Rev.}\ }\textbf {\bibinfo {volume} {111}},\ \bibinfo {pages} {412} (\bibinfo
  {year} {1958})}\BibitemShut {NoStop}%
\bibitem [{\citenamefont {Gurevich}(2017)}]{Gurevich_SUST_2017}%
  \BibitemOpen
  \bibfield  {author} {\bibinfo {author} {\bibfnamefont {A.}~\bibnamefont
  {Gurevich}},\ }\href@noop {} {\bibfield  {journal} {\bibinfo  {journal}
  {Supercond. Sci. Technol.}\ }\textbf {\bibinfo {volume} {30}},\ \bibinfo
  {pages} {034004} (\bibinfo {year} {2017})}\BibitemShut {NoStop}%
\bibitem [{\citenamefont {Kubo}\ and\ \citenamefont
  {Gurevich}(2019)}]{Kubo_PRB_2019}%
  \BibitemOpen
  \bibfield  {author} {\bibinfo {author} {\bibfnamefont {T.}~\bibnamefont
  {Kubo}}\ and\ \bibinfo {author} {\bibfnamefont {A.}~\bibnamefont
  {Gurevich}},\ }\href@noop {} {\bibfield  {journal} {\bibinfo  {journal}
  {Phys. Rev. B}\ }\textbf {\bibinfo {volume} {100}},\ \bibinfo {pages}
  {064522} (\bibinfo {year} {2019})}\BibitemShut {NoStop}%
\bibitem [{\citenamefont {Romanenko}\ \emph
  {et~al.}(2013{\natexlab{a}})\citenamefont {Romanenko}, \citenamefont
  {Barkov}, \citenamefont {Cooley},\ and\ \citenamefont
  {Grassellino}}]{Romanenko_SUST_2013}%
  \BibitemOpen
  \bibfield  {author} {\bibinfo {author} {\bibfnamefont {A.}~\bibnamefont
  {Romanenko}}, \bibinfo {author} {\bibfnamefont {F.}~\bibnamefont {Barkov}},
  \bibinfo {author} {\bibfnamefont {L.~D.}\ \bibnamefont {Cooley}}, \ and\
  \bibinfo {author} {\bibfnamefont {A.}~\bibnamefont {Grassellino}},\
  }\href@noop {} {\bibfield  {journal} {\bibinfo  {journal} {Supercond. Sci.
  Technol.}\ }\textbf {\bibinfo {volume} {26}},\ \bibinfo {pages} {035003}
  (\bibinfo {year} {2013}{\natexlab{a}})}\BibitemShut {NoStop}%
\bibitem [{\citenamefont {Romanenko}\ and\ \citenamefont
  {Schuster}(2017)}]{Romanenko_PRL_2017}%
  \BibitemOpen
  \bibfield  {author} {\bibinfo {author} {\bibfnamefont {A.}~\bibnamefont
  {Romanenko}}\ and\ \bibinfo {author} {\bibfnamefont {D.}~\bibnamefont
  {Schuster}},\ }\href@noop {} {\bibfield  {journal} {\bibinfo  {journal}
  {Phys. Rev. Lett.}\ }\textbf {\bibinfo {volume} {119}},\ \bibinfo {pages}
  {264801} (\bibinfo {year} {2017})}\BibitemShut {NoStop}%
\bibitem [{\citenamefont {Romanenko}\ \emph {et~al.}(2018)\citenamefont
  {Romanenko}, \citenamefont {Pilipenko}, \citenamefont {Zorzetti},
  \citenamefont {Frolov}, \citenamefont {Awida}, \citenamefont {Belomestnykh},
  \citenamefont {Posen},\ and\ \citenamefont
  {Grassellino}}]{Romanenko_arXiv_2018}%
  \BibitemOpen
  \bibfield  {author} {\bibinfo {author} {\bibfnamefont {A.}~\bibnamefont
  {Romanenko}}, \bibinfo {author} {\bibfnamefont {R.}~\bibnamefont
  {Pilipenko}}, \bibinfo {author} {\bibfnamefont {S.}~\bibnamefont {Zorzetti}},
  \bibinfo {author} {\bibfnamefont {D.}~\bibnamefont {Frolov}}, \bibinfo
  {author} {\bibfnamefont {M.}~\bibnamefont {Awida}}, \bibinfo {author}
  {\bibfnamefont {S.}~\bibnamefont {Belomestnykh}}, \bibinfo {author}
  {\bibfnamefont {S.}~\bibnamefont {Posen}}, \ and\ \bibinfo {author}
  {\bibfnamefont {A.}~\bibnamefont {Grassellino}},\ }\href@noop {} {\
  (\bibinfo {year} {2018})},\ \Eprint {http://arxiv.org/abs/1810.03703}
  {arXiv:1810.03703} \BibitemShut {NoStop}%
\bibitem [{\citenamefont {Cardona}\ \emph {et~al.}(1970)\citenamefont
  {Cardona}, \citenamefont {Fischer},\ and\ \citenamefont
  {Rosenblum}}]{Cadorna_PRL_1964}%
  \BibitemOpen
  \bibfield  {author} {\bibinfo {author} {\bibfnamefont {M.}~\bibnamefont
  {Cardona}}, \bibinfo {author} {\bibfnamefont {G.}~\bibnamefont {Fischer}}, \
  and\ \bibinfo {author} {\bibfnamefont {B.}~\bibnamefont {Rosenblum}},\
  }\href@noop {} {\bibfield  {journal} {\bibinfo  {journal} {Phys. Rev. Lett.}\
  }\textbf {\bibinfo {volume} {12}},\ \bibinfo {pages} {101} (\bibinfo {year}
  {1970})}\BibitemShut {NoStop}%
\bibitem [{\citenamefont {Gittleman}\ and\ \citenamefont
  {Rosenblum}(1966)}]{Gittleman_PRL_1966}%
  \BibitemOpen
  \bibfield  {author} {\bibinfo {author} {\bibfnamefont {J.~I.}\ \bibnamefont
  {Gittleman}}\ and\ \bibinfo {author} {\bibfnamefont {B.}~\bibnamefont
  {Rosenblum}},\ }\href@noop {} {\bibfield  {journal} {\bibinfo  {journal}
  {Phys. Rev. Lett.}\ }\textbf {\bibinfo {volume} {16}},\ \bibinfo {pages}
  {734} (\bibinfo {year} {1966})}\BibitemShut {NoStop}%
\bibitem [{\citenamefont {Rabinowitz}(1971)}]{Rabinowitz_JAP_1971}%
  \BibitemOpen
  \bibfield  {author} {\bibinfo {author} {\bibfnamefont {M.}~\bibnamefont
  {Rabinowitz}},\ }\href@noop {} {\bibfield  {journal} {\bibinfo  {journal} {J.
  Appl. Phys.}\ }\textbf {\bibinfo {volume} {42}},\ \bibinfo {pages} {88}
  (\bibinfo {year} {1971})}\BibitemShut {NoStop}%
\bibitem [{\citenamefont {Chin}\ \emph {et~al.}(1992)\citenamefont {Chin},
  \citenamefont {Oates}, \citenamefont {Dresselhaus},\ and\ \citenamefont
  {Dresselhaus}}]{Chin_PRB_1992}%
  \BibitemOpen
  \bibfield  {author} {\bibinfo {author} {\bibfnamefont {C.~C.}\ \bibnamefont
  {Chin}}, \bibinfo {author} {\bibfnamefont {D.~E.}\ \bibnamefont {Oates}},
  \bibinfo {author} {\bibfnamefont {G.}~\bibnamefont {Dresselhaus}}, \ and\
  \bibinfo {author} {\bibfnamefont {M.~S.}\ \bibnamefont {Dresselhaus}},\
  }\href@noop {} {\bibfield  {journal} {\bibinfo  {journal} {Phys. Rev. B}\
  }\textbf {\bibinfo {volume} {45}},\ \bibinfo {pages} {4788} (\bibinfo {year}
  {1992})}\BibitemShut {NoStop}%
\bibitem [{\citenamefont {Janju\v{s}evi\'{c}}\ \emph
  {et~al.}(2006)\citenamefont {Janju\v{s}evi\'{c}}, \citenamefont {Grbi\'{c}},
  \citenamefont {Po\v{z}ek}, \citenamefont {Dul\v{c}i\'{c}}, \citenamefont
  {Paar}, \citenamefont {Nebendahl},\ and\ \citenamefont
  {Wagner}}]{Janjusevic_PRB_2006}%
  \BibitemOpen
  \bibfield  {author} {\bibinfo {author} {\bibfnamefont {D.}~\bibnamefont
  {Janju\v{s}evi\'{c}}}, \bibinfo {author} {\bibfnamefont {M.~S.}\ \bibnamefont
  {Grbi\'{c}}}, \bibinfo {author} {\bibfnamefont {M.}~\bibnamefont
  {Po\v{z}ek}}, \bibinfo {author} {\bibfnamefont {A.}~\bibnamefont
  {Dul\v{c}i\'{c}}}, \bibinfo {author} {\bibfnamefont {D.}~\bibnamefont
  {Paar}}, \bibinfo {author} {\bibfnamefont {B.}~\bibnamefont {Nebendahl}}, \
  and\ \bibinfo {author} {\bibfnamefont {T.}~\bibnamefont {Wagner}},\
  }\href@noop {} {\bibfield  {journal} {\bibinfo  {journal} {Phys. Rev. B}\
  }\textbf {\bibinfo {volume} {74}},\ \bibinfo {pages} {104501} (\bibinfo
  {year} {2006})}\BibitemShut {NoStop}%
\bibitem [{\citenamefont {Alimenti}\ \emph {et~al.}(2019)\citenamefont
  {Alimenti}, \citenamefont {Pompeo}, \citenamefont {Torokhtii}, \citenamefont
  {Spina}, \citenamefont {Fl{\"u}kiger}, \citenamefont {Muzzi},\ and\
  \citenamefont {Silva}}]{Alimenti_IEEE_2019}%
  \BibitemOpen
  \bibfield  {author} {\bibinfo {author} {\bibfnamefont {A.}~\bibnamefont
  {Alimenti}}, \bibinfo {author} {\bibfnamefont {N.}~\bibnamefont {Pompeo}},
  \bibinfo {author} {\bibfnamefont {K.}~\bibnamefont {Torokhtii}}, \bibinfo
  {author} {\bibfnamefont {T.}~\bibnamefont {Spina}}, \bibinfo {author}
  {\bibfnamefont {R.}~\bibnamefont {Fl{\"u}kiger}}, \bibinfo {author}
  {\bibfnamefont {L.}~\bibnamefont {Muzzi}}, \ and\ \bibinfo {author}
  {\bibfnamefont {E.}~\bibnamefont {Silva}},\ }\href@noop {} {\bibfield
  {journal} {\bibinfo  {journal} {IEEE Trans. Appl. Supercond.}\ }\textbf
  {\bibinfo {volume} {29}},\ \bibinfo {pages} {3500104} (\bibinfo {year}
  {2019})}\BibitemShut {NoStop}%
\bibitem [{\citenamefont {Revenaz}\ \emph {et~al.}(1994)\citenamefont
  {Revenaz}, \citenamefont {D.~E.~Oates}, \citenamefont {Dresselhaus},\ and\
  \citenamefont {Dresselhaus}}]{Revenaz_PRB_1994}%
  \BibitemOpen
  \bibfield  {author} {\bibinfo {author} {\bibfnamefont {S.}~\bibnamefont
  {Revenaz}}, \bibinfo {author} {\bibfnamefont {D.~L.-L.}\ \bibnamefont
  {D.~E.~Oates}}, \bibinfo {author} {\bibfnamefont {G.}~\bibnamefont
  {Dresselhaus}}, \ and\ \bibinfo {author} {\bibfnamefont {M.~S.}\ \bibnamefont
  {Dresselhaus}},\ }\href@noop {} {\bibfield  {journal} {\bibinfo  {journal}
  {Phys. Rev. B}\ }\textbf {\bibinfo {volume} {50}},\ \bibinfo {pages} {1178}
  (\bibinfo {year} {1994})}\BibitemShut {NoStop}%
\bibitem [{\citenamefont {Matsuda}\ \emph {et~al.}(1995)\citenamefont
  {Matsuda}, \citenamefont {Gaifullin}, \citenamefont {Kumagai}, \citenamefont
  {Kadowaki},\ and\ \citenamefont {Mochiku}}]{Matsuda_PRL_1995}%
  \BibitemOpen
  \bibfield  {author} {\bibinfo {author} {\bibfnamefont {Y.}~\bibnamefont
  {Matsuda}}, \bibinfo {author} {\bibfnamefont {M.~B.}\ \bibnamefont
  {Gaifullin}}, \bibinfo {author} {\bibfnamefont {K.}~\bibnamefont {Kumagai}},
  \bibinfo {author} {\bibfnamefont {K.}~\bibnamefont {Kadowaki}}, \ and\
  \bibinfo {author} {\bibfnamefont {T.}~\bibnamefont {Mochiku}},\ }\href@noop
  {} {\bibfield  {journal} {\bibinfo  {journal} {Phys. Rev. Lett.}\ }\textbf
  {\bibinfo {volume} {75}},\ \bibinfo {pages} {4512} (\bibinfo {year}
  {1995})}\BibitemShut {NoStop}%
\bibitem [{\citenamefont {Golosovsky}\ \emph {et~al.}(1996)\citenamefont
  {Golosovsky}, \citenamefont {Tsindlekht},\ and\ \citenamefont
  {Davidov}}]{Golosovsky_SUST_1996}%
  \BibitemOpen
  \bibfield  {author} {\bibinfo {author} {\bibfnamefont {M.}~\bibnamefont
  {Golosovsky}}, \bibinfo {author} {\bibfnamefont {M.}~\bibnamefont
  {Tsindlekht}}, \ and\ \bibinfo {author} {\bibfnamefont {D.}~\bibnamefont
  {Davidov}},\ }\href@noop {} {\bibfield  {journal} {\bibinfo  {journal}
  {Supercond. Sci. Technol.}\ }\textbf {\bibinfo {volume} {9}},\ \bibinfo
  {pages} {1} (\bibinfo {year} {1996})}\BibitemShut {NoStop}%
\bibitem [{\citenamefont {Hashimoto}\ \emph
  {et~al.}(2009{\natexlab{a}})\citenamefont {Hashimoto}, \citenamefont
  {Shibauchi}, \citenamefont {Kato}, \citenamefont {Ikada}, \citenamefont
  {Okazaki}, \citenamefont {Shishido}, \citenamefont {Ishikado}, \citenamefont
  {Kito}, \citenamefont {Iyo}, \citenamefont {Eisaki}, \citenamefont
  {Shamoto},\ and\ \citenamefont {Matsuda}}]{Hashimoto_PRL_2009_1}%
  \BibitemOpen
  \bibfield  {author} {\bibinfo {author} {\bibfnamefont {K.}~\bibnamefont
  {Hashimoto}}, \bibinfo {author} {\bibfnamefont {T.}~\bibnamefont
  {Shibauchi}}, \bibinfo {author} {\bibfnamefont {T.}~\bibnamefont {Kato}},
  \bibinfo {author} {\bibfnamefont {K.}~\bibnamefont {Ikada}}, \bibinfo
  {author} {\bibfnamefont {R.}~\bibnamefont {Okazaki}}, \bibinfo {author}
  {\bibfnamefont {H.}~\bibnamefont {Shishido}}, \bibinfo {author}
  {\bibfnamefont {M.}~\bibnamefont {Ishikado}}, \bibinfo {author}
  {\bibfnamefont {H.}~\bibnamefont {Kito}}, \bibinfo {author} {\bibfnamefont
  {A.}~\bibnamefont {Iyo}}, \bibinfo {author} {\bibfnamefont {H.}~\bibnamefont
  {Eisaki}}, \bibinfo {author} {\bibfnamefont {S.}~\bibnamefont {Shamoto}}, \
  and\ \bibinfo {author} {\bibfnamefont {Y.}~\bibnamefont {Matsuda}},\
  }\href@noop {} {\bibfield  {journal} {\bibinfo  {journal} {Phys. Rev. Lett.}\
  }\textbf {\bibinfo {volume} {102}},\ \bibinfo {pages} {017002} (\bibinfo
  {year} {2009}{\natexlab{a}})}\BibitemShut {NoStop}%
\bibitem [{\citenamefont {Hashimoto}\ \emph
  {et~al.}(2009{\natexlab{b}})\citenamefont {Hashimoto}, \citenamefont
  {Shibauchi}, \citenamefont {Kasahara}, \citenamefont {Ikada}, \citenamefont
  {Tonegawa}, \citenamefont {Kato}, \citenamefont {Okazaki}, \citenamefont
  {van~der Beek}, \citenamefont {Konczykowski}, \citenamefont {Takeya},
  \citenamefont {Hirata}, \citenamefont {Terashima},\ and\ \citenamefont
  {Matsuda}}]{Hashimoto_PRL_2009_2}%
  \BibitemOpen
  \bibfield  {author} {\bibinfo {author} {\bibfnamefont {K.}~\bibnamefont
  {Hashimoto}}, \bibinfo {author} {\bibfnamefont {T.}~\bibnamefont
  {Shibauchi}}, \bibinfo {author} {\bibfnamefont {S.}~\bibnamefont {Kasahara}},
  \bibinfo {author} {\bibfnamefont {K.}~\bibnamefont {Ikada}}, \bibinfo
  {author} {\bibfnamefont {S.}~\bibnamefont {Tonegawa}}, \bibinfo {author}
  {\bibfnamefont {T.}~\bibnamefont {Kato}}, \bibinfo {author} {\bibfnamefont
  {R.}~\bibnamefont {Okazaki}}, \bibinfo {author} {\bibfnamefont {C.~J.}\
  \bibnamefont {van~der Beek}}, \bibinfo {author} {\bibfnamefont
  {M.}~\bibnamefont {Konczykowski}}, \bibinfo {author} {\bibfnamefont
  {H.}~\bibnamefont {Takeya}}, \bibinfo {author} {\bibfnamefont
  {K.}~\bibnamefont {Hirata}}, \bibinfo {author} {\bibfnamefont
  {T.}~\bibnamefont {Terashima}}, \ and\ \bibinfo {author} {\bibfnamefont
  {Y.}~\bibnamefont {Matsuda}},\ }\href@noop {} {\bibfield  {journal} {\bibinfo
   {journal} {Phys. Rev. Lett.}\ }\textbf {\bibinfo {volume} {102}},\ \bibinfo
  {pages} {207001} (\bibinfo {year} {2009}{\natexlab{b}})}\BibitemShut
  {NoStop}%
\bibitem [{\citenamefont {Okada}\ \emph {et~al.}(2012)\citenamefont {Okada},
  \citenamefont {Takahashi}, \citenamefont {Imai}, \citenamefont {Kitagawa},
  \citenamefont {Matsubayashi}, \citenamefont {Uwatoko},\ and\ \citenamefont
  {Maeda}}]{Okada_PRB_2012}%
  \BibitemOpen
  \bibfield  {author} {\bibinfo {author} {\bibfnamefont {T.}~\bibnamefont
  {Okada}}, \bibinfo {author} {\bibfnamefont {H.}~\bibnamefont {Takahashi}},
  \bibinfo {author} {\bibfnamefont {Y.}~\bibnamefont {Imai}}, \bibinfo {author}
  {\bibfnamefont {K.}~\bibnamefont {Kitagawa}}, \bibinfo {author}
  {\bibfnamefont {K.}~\bibnamefont {Matsubayashi}}, \bibinfo {author}
  {\bibfnamefont {Y.}~\bibnamefont {Uwatoko}}, \ and\ \bibinfo {author}
  {\bibfnamefont {A.}~\bibnamefont {Maeda}},\ }\href@noop {} {\bibfield
  {journal} {\bibinfo  {journal} {Phys. Rev. B}\ }\textbf {\bibinfo {volume}
  {86}},\ \bibinfo {pages} {064516} (\bibinfo {year} {2012})}\BibitemShut
  {NoStop}%
\bibitem [{\citenamefont {Coffey}\ and\ \citenamefont
  {Clem}(1991)}]{Coffey_PRL_1991}%
  \BibitemOpen
  \bibfield  {author} {\bibinfo {author} {\bibfnamefont {M.~W.}\ \bibnamefont
  {Coffey}}\ and\ \bibinfo {author} {\bibfnamefont {J.~R.}\ \bibnamefont
  {Clem}},\ }\href@noop {} {\bibfield  {journal} {\bibinfo  {journal} {Phys.
  Rev. Lett.}\ }\textbf {\bibinfo {volume} {67}},\ \bibinfo {pages} {386}
  (\bibinfo {year} {1991})}\BibitemShut {NoStop}%
\bibitem [{\citenamefont {Marcon}\ \emph {et~al.}(1991)\citenamefont {Marcon},
  \citenamefont {Fastampa}, \citenamefont {Giura},\ and\ \citenamefont
  {Silva}}]{Marcon_PRB_1991}%
  \BibitemOpen
  \bibfield  {author} {\bibinfo {author} {\bibfnamefont {R.}~\bibnamefont
  {Marcon}}, \bibinfo {author} {\bibfnamefont {R.}~\bibnamefont {Fastampa}},
  \bibinfo {author} {\bibfnamefont {M.}~\bibnamefont {Giura}}, \ and\ \bibinfo
  {author} {\bibfnamefont {E.}~\bibnamefont {Silva}},\ }\href@noop {}
  {\bibfield  {journal} {\bibinfo  {journal} {Phys. Rev. B}\ }\textbf {\bibinfo
  {volume} {43}},\ \bibinfo {pages} {2940} (\bibinfo {year}
  {1991})}\BibitemShut {NoStop}%
\bibitem [{\citenamefont {Sonin}\ \emph {et~al.}(1992)\citenamefont {Sonin},
  \citenamefont {Tagantsev},\ and\ \citenamefont {Traito}}]{Sonin_PRB_1992}%
  \BibitemOpen
  \bibfield  {author} {\bibinfo {author} {\bibfnamefont {E.~B.}\ \bibnamefont
  {Sonin}}, \bibinfo {author} {\bibfnamefont {A.~K.}\ \bibnamefont
  {Tagantsev}}, \ and\ \bibinfo {author} {\bibfnamefont {K.~B.}\ \bibnamefont
  {Traito}},\ }\href@noop {} {\bibfield  {journal} {\bibinfo  {journal} {Phys.
  Rev. B}\ }\textbf {\bibinfo {volume} {43}},\ \bibinfo {pages} {5830}
  (\bibinfo {year} {1992})}\BibitemShut {NoStop}%
\bibitem [{\citenamefont {Grassellino}\ \emph {et~al.}(2013)\citenamefont
  {Grassellino}, \citenamefont {Romanenko}, \citenamefont {Sergatskov},
  \citenamefont {Melnychuk}, \citenamefont {Trenikhina}, \citenamefont
  {Crawford}, \citenamefont {Rowe}, \citenamefont {Wong}, \citenamefont
  {Khabiboulline},\ and\ \citenamefont {Barkov}}]{Grassellino_SUST_2013}%
  \BibitemOpen
  \bibfield  {author} {\bibinfo {author} {\bibfnamefont {A.}~\bibnamefont
  {Grassellino}}, \bibinfo {author} {\bibfnamefont {A.}~\bibnamefont
  {Romanenko}}, \bibinfo {author} {\bibfnamefont {D.~A.}\ \bibnamefont
  {Sergatskov}}, \bibinfo {author} {\bibfnamefont {O.}~\bibnamefont
  {Melnychuk}}, \bibinfo {author} {\bibfnamefont {Y.}~\bibnamefont
  {Trenikhina}}, \bibinfo {author} {\bibfnamefont {A.~C.}\ \bibnamefont
  {Crawford}}, \bibinfo {author} {\bibfnamefont {A.}~\bibnamefont {Rowe}},
  \bibinfo {author} {\bibfnamefont {M.}~\bibnamefont {Wong}}, \bibinfo {author}
  {\bibfnamefont {T.}~\bibnamefont {Khabiboulline}}, \ and\ \bibinfo {author}
  {\bibfnamefont {F.}~\bibnamefont {Barkov}},\ }\href@noop {} {\bibfield
  {journal} {\bibinfo  {journal} {Supercond. Sci. Tech.}\ }\textbf {\bibinfo
  {volume} {26}},\ \bibinfo {pages} {102001} (\bibinfo {year}
  {2013})}\BibitemShut {NoStop}%
\bibitem [{\citenamefont {Martinello}\ \emph {et~al.}(2018)\citenamefont
  {Martinello}, \citenamefont {Checchin}, \citenamefont {Romanenko},
  \citenamefont {Grassellino}, \citenamefont {Aderhold}, \citenamefont
  {Chandrasekeran}, \citenamefont {Melnychuk}, \citenamefont {Posen},\ and\
  \citenamefont {Sergatskov}}]{Martinello_PRL_2018}%
  \BibitemOpen
  \bibfield  {author} {\bibinfo {author} {\bibfnamefont {M.}~\bibnamefont
  {Martinello}}, \bibinfo {author} {\bibfnamefont {M.}~\bibnamefont
  {Checchin}}, \bibinfo {author} {\bibfnamefont {A.}~\bibnamefont {Romanenko}},
  \bibinfo {author} {\bibfnamefont {A.}~\bibnamefont {Grassellino}}, \bibinfo
  {author} {\bibfnamefont {S.}~\bibnamefont {Aderhold}}, \bibinfo {author}
  {\bibfnamefont {S.~K.}\ \bibnamefont {Chandrasekeran}}, \bibinfo {author}
  {\bibfnamefont {O.}~\bibnamefont {Melnychuk}}, \bibinfo {author}
  {\bibfnamefont {S.}~\bibnamefont {Posen}}, \ and\ \bibinfo {author}
  {\bibfnamefont {D.~A.}\ \bibnamefont {Sergatskov}},\ }\href@noop {}
  {\bibfield  {journal} {\bibinfo  {journal} {Phys. Rev. Lett.}\ }\textbf
  {\bibinfo {volume} {121}},\ \bibinfo {pages} {224801} (\bibinfo {year}
  {2018})}\BibitemShut {NoStop}%
\bibitem [{\citenamefont {Piosczyk}\ \emph {et~al.}(1973)\citenamefont
  {Piosczyk}, \citenamefont {Kneisel}, \citenamefont {Stoltz},\ and\
  \citenamefont {Halbritter}}]{Piosczyk_IEEE_1973}%
  \BibitemOpen
  \bibfield  {author} {\bibinfo {author} {\bibfnamefont {B.}~\bibnamefont
  {Piosczyk}}, \bibinfo {author} {\bibfnamefont {P.}~\bibnamefont {Kneisel}},
  \bibinfo {author} {\bibfnamefont {O.}~\bibnamefont {Stoltz}}, \ and\ \bibinfo
  {author} {\bibfnamefont {J.}~\bibnamefont {Halbritter}},\ }\href@noop {}
  {\bibfield  {journal} {\bibinfo  {journal} {IEEE Trans. Nucl. Sci.}\ }\textbf
  {\bibinfo {volume} {20}},\ \bibinfo {pages} {108} (\bibinfo {year}
  {1973})}\BibitemShut {NoStop}%
\bibitem [{\citenamefont {Benvenuti}\ \emph {et~al.}(1999)\citenamefont
  {Benvenuti}, \citenamefont {Calatroni}, \citenamefont {Campisi},
  \citenamefont {Darriulat}, \citenamefont {Peck}, \citenamefont {Russo},\ and\
  \citenamefont {Valente}}]{Benvenuti_PhysicaC_1999}%
  \BibitemOpen
  \bibfield  {author} {\bibinfo {author} {\bibfnamefont {C.}~\bibnamefont
  {Benvenuti}}, \bibinfo {author} {\bibfnamefont {S.}~\bibnamefont
  {Calatroni}}, \bibinfo {author} {\bibfnamefont {I.~E.}\ \bibnamefont
  {Campisi}}, \bibinfo {author} {\bibfnamefont {P.}~\bibnamefont {Darriulat}},
  \bibinfo {author} {\bibfnamefont {M.~A.}\ \bibnamefont {Peck}}, \bibinfo
  {author} {\bibfnamefont {R.}~\bibnamefont {Russo}}, \ and\ \bibinfo {author}
  {\bibfnamefont {A.-M.}\ \bibnamefont {Valente}},\ }\href@noop {} {\bibfield
  {journal} {\bibinfo  {journal} {Physica C}\ }\textbf {\bibinfo {volume}
  {316}},\ \bibinfo {pages} {153} (\bibinfo {year} {1999})}\BibitemShut
  {NoStop}%
\bibitem [{\citenamefont {Gurevich}\ and\ \citenamefont
  {Ciovati}(2013)}]{Gurevich_PRB_2013}%
  \BibitemOpen
  \bibfield  {author} {\bibinfo {author} {\bibfnamefont {A.}~\bibnamefont
  {Gurevich}}\ and\ \bibinfo {author} {\bibfnamefont {G.}~\bibnamefont
  {Ciovati}},\ }\href@noop {} {\bibfield  {journal} {\bibinfo  {journal} {Phys.
  Rev. B}\ }\textbf {\bibinfo {volume} {87}},\ \bibinfo {pages} {054502}
  (\bibinfo {year} {2013})}\BibitemShut {NoStop}%
\bibitem [{\citenamefont {Gonnella}\ \emph {et~al.}(2016)\citenamefont
  {Gonnella}, \citenamefont {Kaufman},\ and\ \citenamefont
  {Liepe}}]{Gonnella_JAP_2016}%
  \BibitemOpen
  \bibfield  {author} {\bibinfo {author} {\bibfnamefont {D.}~\bibnamefont
  {Gonnella}}, \bibinfo {author} {\bibfnamefont {J.}~\bibnamefont {Kaufman}}, \
  and\ \bibinfo {author} {\bibfnamefont {M.}~\bibnamefont {Liepe}},\
  }\href@noop {} {\bibfield  {journal} {\bibinfo  {journal} {J. Appl. Phys}\
  }\textbf {\bibinfo {volume} {119}},\ \bibinfo {pages} {073904} (\bibinfo
  {year} {2016})}\BibitemShut {NoStop}%
\bibitem [{\citenamefont {Checchin}\ \emph {et~al.}(2017)\citenamefont
  {Checchin}, \citenamefont {Martinello}, \citenamefont {Grassellino},
  \citenamefont {Romanenko},\ and\ \citenamefont
  {Zasadzinski}}]{Checchin_SUST_2017}%
  \BibitemOpen
  \bibfield  {author} {\bibinfo {author} {\bibfnamefont {M.}~\bibnamefont
  {Checchin}}, \bibinfo {author} {\bibfnamefont {M.}~\bibnamefont
  {Martinello}}, \bibinfo {author} {\bibfnamefont {A.}~\bibnamefont
  {Grassellino}}, \bibinfo {author} {\bibfnamefont {A.}~\bibnamefont
  {Romanenko}}, \ and\ \bibinfo {author} {\bibfnamefont {J.~F.}\ \bibnamefont
  {Zasadzinski}},\ }\href@noop {} {\bibfield  {journal} {\bibinfo  {journal}
  {Supercond. Sci. Technol.}\ }\textbf {\bibinfo {volume} {30}},\ \bibinfo
  {pages} {034003} (\bibinfo {year} {2017})}\BibitemShut {NoStop}%
\bibitem [{\citenamefont {Checchin}\ \emph {et~al.}(2018)\citenamefont
  {Checchin}, \citenamefont {Martinello}, \citenamefont {Grassellino},
  \citenamefont {Aderhold}, \citenamefont {Chandrasekaran}, \citenamefont
  {Melnychuk}, \citenamefont {Posen}, \citenamefont {Romanenko},\ and\
  \citenamefont {Sergatskov}}]{Checchin_APL_2018}%
  \BibitemOpen
  \bibfield  {author} {\bibinfo {author} {\bibfnamefont {M.}~\bibnamefont
  {Checchin}}, \bibinfo {author} {\bibfnamefont {M.}~\bibnamefont
  {Martinello}}, \bibinfo {author} {\bibfnamefont {A.}~\bibnamefont
  {Grassellino}}, \bibinfo {author} {\bibfnamefont {S.}~\bibnamefont
  {Aderhold}}, \bibinfo {author} {\bibfnamefont {S.~K.}\ \bibnamefont
  {Chandrasekaran}}, \bibinfo {author} {\bibfnamefont {O.~S.}\ \bibnamefont
  {Melnychuk}}, \bibinfo {author} {\bibfnamefont {S.}~\bibnamefont {Posen}},
  \bibinfo {author} {\bibfnamefont {A.}~\bibnamefont {Romanenko}}, \ and\
  \bibinfo {author} {\bibfnamefont {D.~A.}\ \bibnamefont {Sergatskov}},\
  }\href@noop {} {\bibfield  {journal} {\bibinfo  {journal} {Appl. Phys.
  Lett.}\ }\textbf {\bibinfo {volume} {112}},\ \bibinfo {pages} {072601}
  (\bibinfo {year} {2018})}\BibitemShut {NoStop}%
\bibitem [{\citenamefont {Calatroni}\ and\ \citenamefont
  {Vaglio}(2017)}]{Calatroni_IEEE_2017}%
  \BibitemOpen
  \bibfield  {author} {\bibinfo {author} {\bibfnamefont {S.}~\bibnamefont
  {Calatroni}}\ and\ \bibinfo {author} {\bibfnamefont {R.}~\bibnamefont
  {Vaglio}},\ }\href@noop {} {\bibfield  {journal} {\bibinfo  {journal} {IEEE
  Trans. Appl. Supercond.}\ }\textbf {\bibinfo {volume} {27}},\ \bibinfo
  {pages} {3500506} (\bibinfo {year} {2017})}\BibitemShut {NoStop}%
\bibitem [{\citenamefont {Calatroni}\ and\ \citenamefont
  {Vaglio}(2019)}]{Calatroni_PRAB_2019}%
  \BibitemOpen
  \bibfield  {author} {\bibinfo {author} {\bibfnamefont {S.}~\bibnamefont
  {Calatroni}}\ and\ \bibinfo {author} {\bibfnamefont {R.}~\bibnamefont
  {Vaglio}},\ }\href@noop {} {\bibfield  {journal} {\bibinfo  {journal} {Phys.
  Rev. Accel. Beams}\ }\textbf {\bibinfo {volume} {22}},\ \bibinfo {pages}
  {022001} (\bibinfo {year} {2019})}\BibitemShut {NoStop}%
\bibitem [{\citenamefont {Liarte}\ \emph {et~al.}(2018)\citenamefont {Liarte},
  \citenamefont {Hall}, \citenamefont {Koufalis}, \citenamefont {Miyazaki},
  \citenamefont {Senanian}, \citenamefont {Liepe},\ and\ \citenamefont
  {Sethna}}]{Liarte_PRApplied_2018}%
  \BibitemOpen
  \bibfield  {author} {\bibinfo {author} {\bibfnamefont {D.~B.}\ \bibnamefont
  {Liarte}}, \bibinfo {author} {\bibfnamefont {D.}~\bibnamefont {Hall}},
  \bibinfo {author} {\bibfnamefont {P.~N.}\ \bibnamefont {Koufalis}}, \bibinfo
  {author} {\bibfnamefont {A.}~\bibnamefont {Miyazaki}}, \bibinfo {author}
  {\bibfnamefont {A.}~\bibnamefont {Senanian}}, \bibinfo {author}
  {\bibfnamefont {M.}~\bibnamefont {Liepe}}, \ and\ \bibinfo {author}
  {\bibfnamefont {J.~P.}\ \bibnamefont {Sethna}},\ }\href@noop {} {\bibfield
  {journal} {\bibinfo  {journal} {Phys. Rev. Applied}\ }\textbf {\bibinfo
  {volume} {10}},\ \bibinfo {pages} {054057} (\bibinfo {year}
  {2018})}\BibitemShut {NoStop}%
\bibitem [{\citenamefont {Aune}\ \emph {et~al.}(2000)\citenamefont {Aune},
  \citenamefont {Bandelmann}, \citenamefont {Bloess}, \citenamefont {Bonin},
  \citenamefont {Bosotti}, \citenamefont {Champion}, \citenamefont {Crawford},
  \citenamefont {Deppe}, \citenamefont {Dwersteg}, \citenamefont {Edwards},
  \citenamefont {Edwards}, \citenamefont {Ferrario}, \citenamefont {Fouaidy},
  \citenamefont {Gall}, \citenamefont {Gamp}, \citenamefont {G\"ossel},
  \citenamefont {Graber}, \citenamefont {Hubert}, \citenamefont {H\"uning},
  \citenamefont {Juillard}, \citenamefont {Junquera}, \citenamefont {Kaiser},
  \citenamefont {Kreps}, \citenamefont {Kuchnir}, \citenamefont {Lange},
  \citenamefont {Leenen}, \citenamefont {Liepe}, \citenamefont {Lilje},
  \citenamefont {Matheisen}, \citenamefont {M\"oller}, \citenamefont {Mosnier},
  \citenamefont {Padamsee}, \citenamefont {Pagani}, \citenamefont {Pekeler},
  \citenamefont {Peters}, \citenamefont {Peters}, \citenamefont {Proch},
  \citenamefont {Rehlich}, \citenamefont {Reschke}, \citenamefont {Safa},
  \citenamefont {Schilcher}, \citenamefont {Schm\"user}, \citenamefont
  {Sekutowicz}, \citenamefont {Simrock}, \citenamefont {Singer}, \citenamefont
  {Tigner}, \citenamefont {Trines}, \citenamefont {Twarowski}, \citenamefont
  {Weichert}, \citenamefont {Weisend}, \citenamefont {Wojtkiewicz},
  \citenamefont {Wolff},\ and\ \citenamefont
  {Zapfe}}]{TESLA_Cavities_PRST_2000}%
  \BibitemOpen
  \bibfield  {author} {\bibinfo {author} {\bibfnamefont {B.}~\bibnamefont
  {Aune}}, \bibinfo {author} {\bibfnamefont {R.}~\bibnamefont {Bandelmann}},
  \bibinfo {author} {\bibfnamefont {D.}~\bibnamefont {Bloess}}, \bibinfo
  {author} {\bibfnamefont {B.}~\bibnamefont {Bonin}}, \bibinfo {author}
  {\bibfnamefont {A.}~\bibnamefont {Bosotti}}, \bibinfo {author} {\bibfnamefont
  {M.}~\bibnamefont {Champion}}, \bibinfo {author} {\bibfnamefont
  {C.}~\bibnamefont {Crawford}}, \bibinfo {author} {\bibfnamefont
  {G.}~\bibnamefont {Deppe}}, \bibinfo {author} {\bibfnamefont
  {B.}~\bibnamefont {Dwersteg}}, \bibinfo {author} {\bibfnamefont
  {D.}~\bibnamefont {Edwards}}, \bibinfo {author} {\bibfnamefont {H.~T.}\
  \bibnamefont {Edwards}}, \bibinfo {author} {\bibfnamefont {M.}~\bibnamefont
  {Ferrario}}, \bibinfo {author} {\bibfnamefont {M.}~\bibnamefont {Fouaidy}},
  \bibinfo {author} {\bibfnamefont {P.-D.}\ \bibnamefont {Gall}}, \bibinfo
  {author} {\bibfnamefont {A.}~\bibnamefont {Gamp}}, \bibinfo {author}
  {\bibfnamefont {A.}~\bibnamefont {G\"ossel}}, \bibinfo {author}
  {\bibfnamefont {J.}~\bibnamefont {Graber}}, \bibinfo {author} {\bibfnamefont
  {D.}~\bibnamefont {Hubert}}, \bibinfo {author} {\bibfnamefont
  {M.}~\bibnamefont {H\"uning}}, \bibinfo {author} {\bibfnamefont
  {M.}~\bibnamefont {Juillard}}, \bibinfo {author} {\bibfnamefont
  {T.}~\bibnamefont {Junquera}}, \bibinfo {author} {\bibfnamefont
  {H.}~\bibnamefont {Kaiser}}, \bibinfo {author} {\bibfnamefont
  {G.}~\bibnamefont {Kreps}}, \bibinfo {author} {\bibfnamefont
  {M.}~\bibnamefont {Kuchnir}}, \bibinfo {author} {\bibfnamefont
  {R.}~\bibnamefont {Lange}}, \bibinfo {author} {\bibfnamefont
  {M.}~\bibnamefont {Leenen}}, \bibinfo {author} {\bibfnamefont
  {M.}~\bibnamefont {Liepe}}, \bibinfo {author} {\bibfnamefont
  {L.}~\bibnamefont {Lilje}}, \bibinfo {author} {\bibfnamefont
  {A.}~\bibnamefont {Matheisen}}, \bibinfo {author} {\bibfnamefont {W.-D.}\
  \bibnamefont {M\"oller}}, \bibinfo {author} {\bibfnamefont {A.}~\bibnamefont
  {Mosnier}}, \bibinfo {author} {\bibfnamefont {H.}~\bibnamefont {Padamsee}},
  \bibinfo {author} {\bibfnamefont {C.}~\bibnamefont {Pagani}}, \bibinfo
  {author} {\bibfnamefont {M.}~\bibnamefont {Pekeler}}, \bibinfo {author}
  {\bibfnamefont {H.-B.}\ \bibnamefont {Peters}}, \bibinfo {author}
  {\bibfnamefont {O.}~\bibnamefont {Peters}}, \bibinfo {author} {\bibfnamefont
  {D.}~\bibnamefont {Proch}}, \bibinfo {author} {\bibfnamefont
  {K.}~\bibnamefont {Rehlich}}, \bibinfo {author} {\bibfnamefont
  {D.}~\bibnamefont {Reschke}}, \bibinfo {author} {\bibfnamefont
  {H.}~\bibnamefont {Safa}}, \bibinfo {author} {\bibfnamefont {T.}~\bibnamefont
  {Schilcher}}, \bibinfo {author} {\bibfnamefont {P.}~\bibnamefont
  {Schm\"user}}, \bibinfo {author} {\bibfnamefont {J.}~\bibnamefont
  {Sekutowicz}}, \bibinfo {author} {\bibfnamefont {S.}~\bibnamefont {Simrock}},
  \bibinfo {author} {\bibfnamefont {W.}~\bibnamefont {Singer}}, \bibinfo
  {author} {\bibfnamefont {M.}~\bibnamefont {Tigner}}, \bibinfo {author}
  {\bibfnamefont {D.}~\bibnamefont {Trines}}, \bibinfo {author} {\bibfnamefont
  {K.}~\bibnamefont {Twarowski}}, \bibinfo {author} {\bibfnamefont
  {G.}~\bibnamefont {Weichert}}, \bibinfo {author} {\bibfnamefont
  {J.}~\bibnamefont {Weisend}}, \bibinfo {author} {\bibfnamefont
  {J.}~\bibnamefont {Wojtkiewicz}}, \bibinfo {author} {\bibfnamefont
  {S.}~\bibnamefont {Wolff}}, \ and\ \bibinfo {author} {\bibfnamefont
  {K.}~\bibnamefont {Zapfe}},\ }\href@noop {} {\bibfield  {journal} {\bibinfo
  {journal} {Phys. Rev. ST Accel. Beams}\ }\textbf {\bibinfo {volume} {3}},\
  \bibinfo {pages} {092001} (\bibinfo {year} {2000})}\BibitemShut {NoStop}%
\bibitem [{\citenamefont {Knobloch}\ \emph {et~al.}(1994)\citenamefont
  {Knobloch}, \citenamefont {Muller},\ and\ \citenamefont
  {Padamsee}}]{Knobloch_T_Map_RSI_1994}%
  \BibitemOpen
  \bibfield  {author} {\bibinfo {author} {\bibfnamefont {J.}~\bibnamefont
  {Knobloch}}, \bibinfo {author} {\bibfnamefont {H.}~\bibnamefont {Muller}}, \
  and\ \bibinfo {author} {\bibfnamefont {H.}~\bibnamefont {Padamsee}},\
  }\href@noop {} {\bibfield  {journal} {\bibinfo  {journal} {Rev. Sci.
  Instrum.}\ }\textbf {\bibinfo {volume} {65}},\ \bibinfo {pages} {3521}
  (\bibinfo {year} {1994})}\BibitemShut {NoStop}%
\bibitem [{\citenamefont {Melnychuk}\ \emph {et~al.}(2014)\citenamefont
  {Melnychuk}, \citenamefont {Grassellino},\ and\ \citenamefont
  {Romanenko}}]{Melnychuk_RevSciInstr_2014}%
  \BibitemOpen
  \bibfield  {author} {\bibinfo {author} {\bibfnamefont {O.}~\bibnamefont
  {Melnychuk}}, \bibinfo {author} {\bibfnamefont {A.}~\bibnamefont
  {Grassellino}}, \ and\ \bibinfo {author} {\bibfnamefont {A.}~\bibnamefont
  {Romanenko}},\ }\href@noop {} {\bibfield  {journal} {\bibinfo  {journal}
  {Rev. Sci. Instrum.}\ }\textbf {\bibinfo {volume} {85}},\ \bibinfo {pages}
  {124705} (\bibinfo {year} {2014})}\BibitemShut {NoStop}%
\bibitem [{\citenamefont {London}\ and\ \citenamefont
  {London}(1935)}]{London_ProcRSocLond_1935}%
  \BibitemOpen
  \bibfield  {author} {\bibinfo {author} {\bibfnamefont {F.}~\bibnamefont
  {London}}\ and\ \bibinfo {author} {\bibfnamefont {H.}~\bibnamefont
  {London}},\ }\href@noop {} {\bibfield  {journal} {\bibinfo  {journal} {Proc.
  R. Soc. Lond. A}\ }\textbf {\bibinfo {volume} {149}},\ \bibinfo {pages} {71}
  (\bibinfo {year} {1935})}\BibitemShut {NoStop}%
\bibitem [{\citenamefont {Brandt}(1995)}]{Brandt_RepProgPhys_1995}%
  \BibitemOpen
  \bibfield  {author} {\bibinfo {author} {\bibfnamefont {E.~H.}\ \bibnamefont
  {Brandt}},\ }\href@noop {} {\bibfield  {journal} {\bibinfo  {journal} {Rep.
  Prog. Phys.}\ }\textbf {\bibinfo {volume} {58}},\ \bibinfo {pages} {1465}
  (\bibinfo {year} {1995})}\BibitemShut {NoStop}%
\bibitem [{\citenamefont {Bardeen}\ and\ \citenamefont
  {Stephen}(1965)}]{Bardeen_PR_1965}%
  \BibitemOpen
  \bibfield  {author} {\bibinfo {author} {\bibfnamefont {J.}~\bibnamefont
  {Bardeen}}\ and\ \bibinfo {author} {\bibfnamefont {M.~J.}\ \bibnamefont
  {Stephen}},\ }\href@noop {} {\bibfield  {journal} {\bibinfo  {journal} {Phys.
  Rev.}\ }\textbf {\bibinfo {volume} {140}},\ \bibinfo {pages} {A1197}
  (\bibinfo {year} {1965})}\BibitemShut {NoStop}%
\bibitem [{\citenamefont {Ginzburg}\ and\ \citenamefont
  {Landau}(1950)}]{Ginzburg_ZETF_1950}%
  \BibitemOpen
  \bibfield  {author} {\bibinfo {author} {\bibfnamefont {V.~L.}\ \bibnamefont
  {Ginzburg}}\ and\ \bibinfo {author} {\bibfnamefont {L.~D.}\ \bibnamefont
  {Landau}},\ }\href@noop {} {\bibfield  {journal} {\bibinfo  {journal} {Zh.
  Eksp. Teor. Fiz.}\ }\textbf {\bibinfo {volume} {20}},\ \bibinfo {pages}
  {1064} (\bibinfo {year} {1950})},\ \translation{L. D. Landau, \emph{Collected
  papers} (Pergamon Press, New York, 1965)}\BibitemShut {NoStop}%
\bibitem [{\citenamefont {Martinello}(2019)}]{Martinello_SRF_2019}%
  \BibitemOpen
  \bibfield  {author} {\bibinfo {author} {\bibfnamefont {M.}~\bibnamefont
  {Martinello}},\ }\href
  {http://accelconf.web.cern.ch/AccelConf/srf2019/talks/tufub4_talk.pdf}
  {\enquote {\bibinfo {title} {Microscopic investigation of flux trapping sites
  in bulk nb},}\ } (\bibinfo {year} {2019}),\ \bibinfo {note} {\textit{19th
  International Conference on RF Superconductivity}}\BibitemShut {NoStop}%
\bibitem [{\citenamefont {Antoine}(2019)}]{Antoine_PRAB_2019}%
  \BibitemOpen
  \bibfield  {author} {\bibinfo {author} {\bibfnamefont {C.~Z.}\ \bibnamefont
  {Antoine}},\ }\href@noop {} {\bibfield  {journal} {\bibinfo  {journal} {Phys.
  Rev. Accel. Beams}\ }\textbf {\bibinfo {volume} {22}},\ \bibinfo {pages}
  {034801} (\bibinfo {year} {2019})}\BibitemShut {NoStop}%
\bibitem [{\citenamefont {Thuneberg}\ \emph {et~al.}(1984)\citenamefont
  {Thuneberg}, \citenamefont {J.Kurkij{\"a}rvi},\ and\ \citenamefont
  {Rainer}}]{Thuneberg_PRB_1984}%
  \BibitemOpen
  \bibfield  {author} {\bibinfo {author} {\bibfnamefont {E.~V.}\ \bibnamefont
  {Thuneberg}}, \bibinfo {author} {\bibnamefont {J.Kurkij{\"a}rvi}}, \ and\
  \bibinfo {author} {\bibfnamefont {D.}~\bibnamefont {Rainer}},\ }\href@noop {}
  {\bibfield  {journal} {\bibinfo  {journal} {Phys. Rev. B}\ }\textbf {\bibinfo
  {volume} {29}},\ \bibinfo {pages} {3913} (\bibinfo {year}
  {1984})}\BibitemShut {NoStop}%
\bibitem [{\citenamefont {Embon}\ \emph {et~al.}(2015)\citenamefont {Embon},
  \citenamefont {Anahory}, \citenamefont {Suhov}, \citenamefont {Halbertal},
  \citenamefont {Cuppens}, \citenamefont {Yakovenko}, \citenamefont {Uri},
  \citenamefont {Myasoedov}, \citenamefont {Rappaport}, \citenamefont {Huber},
  \citenamefont {Gurevich},\ and\ \citenamefont {Zeldov}}]{Embon_SRep_2015}%
  \BibitemOpen
  \bibfield  {author} {\bibinfo {author} {\bibfnamefont {L.}~\bibnamefont
  {Embon}}, \bibinfo {author} {\bibfnamefont {Y.}~\bibnamefont {Anahory}},
  \bibinfo {author} {\bibfnamefont {A.}~\bibnamefont {Suhov}}, \bibinfo
  {author} {\bibfnamefont {D.}~\bibnamefont {Halbertal}}, \bibinfo {author}
  {\bibfnamefont {J.}~\bibnamefont {Cuppens}}, \bibinfo {author} {\bibfnamefont
  {A.}~\bibnamefont {Yakovenko}}, \bibinfo {author} {\bibfnamefont
  {A.}~\bibnamefont {Uri}}, \bibinfo {author} {\bibfnamefont {Y.}~\bibnamefont
  {Myasoedov}}, \bibinfo {author} {\bibfnamefont {M.~L.}\ \bibnamefont
  {Rappaport}}, \bibinfo {author} {\bibfnamefont {M.~E.}\ \bibnamefont
  {Huber}}, \bibinfo {author} {\bibfnamefont {A.}~\bibnamefont {Gurevich}}, \
  and\ \bibinfo {author} {\bibfnamefont {E.}~\bibnamefont {Zeldov}},\
  }\href@noop {} {\bibfield  {journal} {\bibinfo  {journal} {Sci. Rep.}\
  }\textbf {\bibinfo {volume} {5}},\ \bibinfo {pages} {7598} (\bibinfo {year}
  {2015})}\BibitemShut {NoStop}%
\bibitem [{\citenamefont {Allen}\ and\ \citenamefont
  {Claassen}(1989)}]{Allen_PRB_1989}%
  \BibitemOpen
  \bibfield  {author} {\bibinfo {author} {\bibfnamefont {L.~H.}\ \bibnamefont
  {Allen}}\ and\ \bibinfo {author} {\bibfnamefont {J.~H.}\ \bibnamefont
  {Claassen}},\ }\href@noop {} {\bibfield  {journal} {\bibinfo  {journal}
  {Phys. Rev. B}\ }\textbf {\bibinfo {volume} {39}},\ \bibinfo {pages} {2054}
  (\bibinfo {year} {1989})}\BibitemShut {NoStop}%
\bibitem [{\citenamefont {Park}\ \emph {et~al.}(1992)\citenamefont {Park},
  \citenamefont {Cunningham}, \citenamefont {Cabrera},\ and\ \citenamefont
  {Huber}}]{Park_PRL_1992}%
  \BibitemOpen
  \bibfield  {author} {\bibinfo {author} {\bibfnamefont {G.~S.}\ \bibnamefont
  {Park}}, \bibinfo {author} {\bibfnamefont {C.~E.}\ \bibnamefont
  {Cunningham}}, \bibinfo {author} {\bibfnamefont {B.}~\bibnamefont {Cabrera}},
  \ and\ \bibinfo {author} {\bibfnamefont {M.~E.}\ \bibnamefont {Huber}},\
  }\href@noop {} {\bibfield  {journal} {\bibinfo  {journal} {Phys. Rev. Lett.}\
  }\textbf {\bibinfo {volume} {68}},\ \bibinfo {pages} {1920} (\bibinfo {year}
  {1992})}\BibitemShut {NoStop}%
\bibitem [{\citenamefont {Romanenko}\ \emph
  {et~al.}(2014{\natexlab{c}})\citenamefont {Romanenko}, \citenamefont
  {Grassellino}, \citenamefont {Barkov}, \citenamefont {Suter}, \citenamefont
  {Salman},\ and\ \citenamefont {Prokscha}}]{Romanenko_APL_2014}%
  \BibitemOpen
  \bibfield  {author} {\bibinfo {author} {\bibfnamefont {A.}~\bibnamefont
  {Romanenko}}, \bibinfo {author} {\bibfnamefont {A.}~\bibnamefont
  {Grassellino}}, \bibinfo {author} {\bibfnamefont {F.}~\bibnamefont {Barkov}},
  \bibinfo {author} {\bibfnamefont {A.}~\bibnamefont {Suter}}, \bibinfo
  {author} {\bibfnamefont {Z.}~\bibnamefont {Salman}}, \ and\ \bibinfo {author}
  {\bibfnamefont {T.}~\bibnamefont {Prokscha}},\ }\href@noop {} {\bibfield
  {journal} {\bibinfo  {journal} {Appl. Phys. Lett.}\ }\textbf {\bibinfo
  {volume} {104}},\ \bibinfo {pages} {072601} (\bibinfo {year}
  {2014}{\natexlab{c}})}\BibitemShut {NoStop}%
\bibitem [{\citenamefont {Larkin}\ and\ \citenamefont
  {Ovchinnikov}(1975)}]{Larkin_JETP_1976}%
  \BibitemOpen
  \bibfield  {author} {\bibinfo {author} {\bibfnamefont {A.~I.}\ \bibnamefont
  {Larkin}}\ and\ \bibinfo {author} {\bibfnamefont {Y.~N.}\ \bibnamefont
  {Ovchinnikov}},\ }\href@noop {} {\bibfield  {journal} {\bibinfo  {journal}
  {Zh. Eksp. Teor. Fiz.}\ }\textbf {\bibinfo {volume} {68}},\ \bibinfo {pages}
  {1915} (\bibinfo {year} {1975})},\ \translation{A. I. Larkin and Yu. N.
  Ovchinnikov, \emph{Sov. Phys.-JETP}, \textbf{41}, 690 (1976)}\BibitemShut
  {NoStop}%
\bibitem [{\citenamefont {Grimaldi}\ \emph {et~al.}(2010)\citenamefont
  {Grimaldi}, \citenamefont {Leo}, \citenamefont {Zola}, \citenamefont {Nigro},
  \citenamefont {Pace}, \citenamefont {Laviano},\ and\ \citenamefont
  {Mezzetti}}]{Grimaldi_PRB_2010}%
  \BibitemOpen
  \bibfield  {author} {\bibinfo {author} {\bibfnamefont {G.}~\bibnamefont
  {Grimaldi}}, \bibinfo {author} {\bibfnamefont {A.}~\bibnamefont {Leo}},
  \bibinfo {author} {\bibfnamefont {D.}~\bibnamefont {Zola}}, \bibinfo {author}
  {\bibfnamefont {A.}~\bibnamefont {Nigro}}, \bibinfo {author} {\bibfnamefont
  {S.}~\bibnamefont {Pace}}, \bibinfo {author} {\bibfnamefont {F.}~\bibnamefont
  {Laviano}}, \ and\ \bibinfo {author} {\bibfnamefont {E.}~\bibnamefont
  {Mezzetti}},\ }\href@noop {} {\bibfield  {journal} {\bibinfo  {journal}
  {Phys. rev. B}\ }\textbf {\bibinfo {volume} {82}},\ \bibinfo {pages} {024512}
  (\bibinfo {year} {2010})}\BibitemShut {NoStop}%
\bibitem [{\citenamefont {Leo}\ \emph {et~al.}(2011)\citenamefont {Leo},
  \citenamefont {Grimaldi}, \citenamefont {Citro}, \citenamefont {Nigro},
  \citenamefont {Pace},\ and\ \citenamefont {Hubener}}]{Leo_PRB_2011}%
  \BibitemOpen
  \bibfield  {author} {\bibinfo {author} {\bibfnamefont {A.}~\bibnamefont
  {Leo}}, \bibinfo {author} {\bibfnamefont {G.}~\bibnamefont {Grimaldi}},
  \bibinfo {author} {\bibfnamefont {R.}~\bibnamefont {Citro}}, \bibinfo
  {author} {\bibfnamefont {A.}~\bibnamefont {Nigro}}, \bibinfo {author}
  {\bibfnamefont {S.}~\bibnamefont {Pace}}, \ and\ \bibinfo {author}
  {\bibfnamefont {R.~P.}\ \bibnamefont {Hubener}},\ }\href@noop {} {\bibfield
  {journal} {\bibinfo  {journal} {Phys. rev. B}\ }\textbf {\bibinfo {volume}
  {84}},\ \bibinfo {pages} {014536} (\bibinfo {year} {2011})}\BibitemShut
  {NoStop}%
\bibitem [{\citenamefont {Dobrovolskiy}\ \emph {et~al.}(2017)\citenamefont
  {Dobrovolskiy}, \citenamefont {Shklovskij}, \citenamefont {Hanefeld},
  \citenamefont {Z{\"o}rb}, \citenamefont {K{\"o}hs},\ and\ \citenamefont
  {Huth}}]{Dobrovolskiy_SUST_2017}%
  \BibitemOpen
  \bibfield  {author} {\bibinfo {author} {\bibfnamefont {O.~V.}\ \bibnamefont
  {Dobrovolskiy}}, \bibinfo {author} {\bibfnamefont {V.~A.}\ \bibnamefont
  {Shklovskij}}, \bibinfo {author} {\bibfnamefont {M.}~\bibnamefont
  {Hanefeld}}, \bibinfo {author} {\bibfnamefont {M.}~\bibnamefont {Z{\"o}rb}},
  \bibinfo {author} {\bibfnamefont {L.}~\bibnamefont {K{\"o}hs}}, \ and\
  \bibinfo {author} {\bibfnamefont {M.}~\bibnamefont {Huth}},\ }\href@noop {}
  {\bibfield  {journal} {\bibinfo  {journal} {Supercond. Sci. Technol.}\
  }\textbf {\bibinfo {volume} {30}},\ \bibinfo {pages} {085002} (\bibinfo
  {year} {2017})}\BibitemShut {NoStop}%
\bibitem [{\citenamefont {Kaplan}\ \emph {et~al.}(1976)\citenamefont {Kaplan},
  \citenamefont {Chi}, \citenamefont {Langenberg}, \citenamefont {Chang},
  \citenamefont {Jafarey},\ and\ \citenamefont {Scalapino}}]{Kaplan_PRB_1976}%
  \BibitemOpen
  \bibfield  {author} {\bibinfo {author} {\bibfnamefont {S.~B.}\ \bibnamefont
  {Kaplan}}, \bibinfo {author} {\bibfnamefont {C.~C.}\ \bibnamefont {Chi}},
  \bibinfo {author} {\bibfnamefont {D.~N.}\ \bibnamefont {Langenberg}},
  \bibinfo {author} {\bibfnamefont {J.~J.}\ \bibnamefont {Chang}}, \bibinfo
  {author} {\bibfnamefont {S.}~\bibnamefont {Jafarey}}, \ and\ \bibinfo
  {author} {\bibfnamefont {D.~J.}\ \bibnamefont {Scalapino}},\ }\href@noop {}
  {\bibfield  {journal} {\bibinfo  {journal} {Phys. Rev. B}\ }\textbf {\bibinfo
  {volume} {14}},\ \bibinfo {pages} {4854} (\bibinfo {year}
  {1976})}\BibitemShut {NoStop}%
\bibitem [{\citenamefont {Ciovati}(2004)}]{Ciovati_JAppPhys_2004}%
  \BibitemOpen
  \bibfield  {author} {\bibinfo {author} {\bibfnamefont {G.}~\bibnamefont
  {Ciovati}},\ }\href@noop {} {\bibfield  {journal} {\bibinfo  {journal} {J.
  Appl. Phys.}\ }\textbf {\bibinfo {volume} {96}},\ \bibinfo {pages} {1591}
  (\bibinfo {year} {2004})}\BibitemShut {NoStop}%
\bibitem [{\citenamefont {Romanenko}\ \emph
  {et~al.}(2013{\natexlab{b}})\citenamefont {Romanenko}, \citenamefont
  {Edwardson}, \citenamefont {Coleman},\ and\ \citenamefont
  {Simpson}}]{Romanenko_APL_2013}%
  \BibitemOpen
  \bibfield  {author} {\bibinfo {author} {\bibfnamefont {A.}~\bibnamefont
  {Romanenko}}, \bibinfo {author} {\bibfnamefont {C.~J.}\ \bibnamefont
  {Edwardson}}, \bibinfo {author} {\bibfnamefont {P.~G.}\ \bibnamefont
  {Coleman}}, \ and\ \bibinfo {author} {\bibfnamefont {P.~J.}\ \bibnamefont
  {Simpson}},\ }\href@noop {} {\bibfield  {journal} {\bibinfo  {journal} {App.
  Phys. Lett.}\ }\textbf {\bibinfo {volume} {102}},\ \bibinfo {pages} {232601}
  (\bibinfo {year} {2013}{\natexlab{b}})}\BibitemShut {NoStop}%
\bibitem [{\citenamefont {Romanenko}\ \emph {et~al.}(2016)\citenamefont
  {Romanenko}, \citenamefont {Trenikhina}, \citenamefont {Martinello},
  \citenamefont {Bafia},\ and\ \citenamefont
  {Grassellino}}]{Romanenko_SRF_2019}%
  \BibitemOpen
  \bibfield  {author} {\bibinfo {author} {\bibfnamefont {A.}~\bibnamefont
  {Romanenko}}, \bibinfo {author} {\bibfnamefont {Y.}~\bibnamefont
  {Trenikhina}}, \bibinfo {author} {\bibfnamefont {M.}~\bibnamefont
  {Martinello}}, \bibinfo {author} {\bibfnamefont {D.}~\bibnamefont {Bafia}}, \
  and\ \bibinfo {author} {\bibfnamefont {A.}~\bibnamefont {Grassellino}},\ }in\
  \href@noop {} {\emph {\bibinfo {booktitle} {Proc. of the 19th International
  Conference on RF Superconductivity}}},\ \bibinfo {series and number}
  {\bibinfo {number} {THP014}}\ (\bibinfo {year} {2016})\ p.\ \bibinfo {pages}
  {866}\BibitemShut {NoStop}%
\bibitem [{\citenamefont {Pippard}(1953)}]{Pippard_ProcRSocLond_1953}%
  \BibitemOpen
  \bibfield  {author} {\bibinfo {author} {\bibfnamefont {A.~B.}\ \bibnamefont
  {Pippard}},\ }\href@noop {} {\bibfield  {journal} {\bibinfo  {journal} {Proc.
  R. Soc. Lond. A}\ }\textbf {\bibinfo {volume} {216}},\ \bibinfo {pages} {547}
  (\bibinfo {year} {1953})}\BibitemShut {NoStop}%
\bibitem [{\citenamefont {Tinkham}(2004)}]{Tinkham_Book}%
  \BibitemOpen
  \bibfield  {author} {\bibinfo {author} {\bibfnamefont {M.}~\bibnamefont
  {Tinkham}},\ }\href@noop {} {\emph {\bibinfo {title} {{Introduction to
  Superconductivity}}}}\ (\bibinfo  {publisher} {Dover Publications},\ \bibinfo
  {year} {2004})\BibitemShut {NoStop}%
\bibitem [{\citenamefont {Bardeen}\ \emph {et~al.}(1957)\citenamefont
  {Bardeen}, \citenamefont {Cooper},\ and\ \citenamefont
  {Schrieffer}}]{Bardeen_PR_1957}%
  \BibitemOpen
  \bibfield  {author} {\bibinfo {author} {\bibfnamefont {J.}~\bibnamefont
  {Bardeen}}, \bibinfo {author} {\bibfnamefont {L.~N.}\ \bibnamefont {Cooper}},
  \ and\ \bibinfo {author} {\bibfnamefont {J.~R.}\ \bibnamefont {Schrieffer}},\
  }\href@noop {} {\bibfield  {journal} {\bibinfo  {journal} {Phys. Rev.}\
  }\textbf {\bibinfo {volume} {108}},\ \bibinfo {pages} {1175} (\bibinfo {year}
  {1957})}\BibitemShut {NoStop}%
\bibitem [{\citenamefont {Gorter}\ and\ \citenamefont
  {Casimir}(1934)}]{Gorter_Physica_1934}%
  \BibitemOpen
  \bibfield  {author} {\bibinfo {author} {\bibfnamefont {C.~J.}\ \bibnamefont
  {Gorter}}\ and\ \bibinfo {author} {\bibfnamefont {H.}~\bibnamefont
  {Casimir}},\ }\href@noop {} {\bibfield  {journal} {\bibinfo  {journal}
  {Physica}\ }\textbf {\bibinfo {volume} {1}},\ \bibinfo {pages} {306}
  (\bibinfo {year} {1934})}\BibitemShut {NoStop}%
\bibitem [{\citenamefont {Suter}\ \emph {et~al.}(2005)\citenamefont {Suter},
  \citenamefont {Morenzoni}, \citenamefont {Garifianov}, \citenamefont
  {Khasanov}, \citenamefont {Kirk}, \citenamefont {Luetkens}, \citenamefont
  {Prokscha},\ and\ \citenamefont {Horisberger}}]{Suter_PRB_2005}%
  \BibitemOpen
  \bibfield  {author} {\bibinfo {author} {\bibfnamefont {A.}~\bibnamefont
  {Suter}}, \bibinfo {author} {\bibfnamefont {E.}~\bibnamefont {Morenzoni}},
  \bibinfo {author} {\bibfnamefont {N.}~\bibnamefont {Garifianov}}, \bibinfo
  {author} {\bibfnamefont {R.}~\bibnamefont {Khasanov}}, \bibinfo {author}
  {\bibfnamefont {E.}~\bibnamefont {Kirk}}, \bibinfo {author} {\bibfnamefont
  {H.}~\bibnamefont {Luetkens}}, \bibinfo {author} {\bibfnamefont
  {T.}~\bibnamefont {Prokscha}}, \ and\ \bibinfo {author} {\bibfnamefont
  {M.}~\bibnamefont {Horisberger}},\ }\href@noop {} {\bibfield  {journal}
  {\bibinfo  {journal} {Phys. Rev. B}\ }\textbf {\bibinfo {volume} {72}},\
  \bibinfo {pages} {024506} (\bibinfo {year} {2005})}\BibitemShut {NoStop}%
\bibitem [{\citenamefont {Sheehan}(1966)}]{Sheehan_PR_1966}%
  \BibitemOpen
  \bibfield  {author} {\bibinfo {author} {\bibfnamefont {T.~P.}\ \bibnamefont
  {Sheehan}},\ }\href@noop {} {\bibfield  {journal} {\bibinfo  {journal} {Phys.
  Rev.}\ }\textbf {\bibinfo {volume} {149}},\ \bibinfo {pages} {368} (\bibinfo
  {year} {1966})}\BibitemShut {NoStop}%
\bibitem [{\citenamefont {Maxfield}\ and\ \citenamefont
  {McLean}(1965)}]{Maxfield_PhysRev_1965}%
  \BibitemOpen
  \bibfield  {author} {\bibinfo {author} {\bibfnamefont {B.~W.}\ \bibnamefont
  {Maxfield}}\ and\ \bibinfo {author} {\bibfnamefont {W.~L.}\ \bibnamefont
  {McLean}},\ }\href@noop {} {\bibfield  {journal} {\bibinfo  {journal} {Phys.
  Rev.}\ }\textbf {\bibinfo {volume} {139}},\ \bibinfo {pages} {A1515}
  (\bibinfo {year} {1965})}\BibitemShut {NoStop}%
\end{thebibliography}%

\end{document}